\documentstyle[12pt]{article}
\makeatletter

\@addtoreset{equation}{section}
\def\theequation{\arabic{section}.\arabic{equation}}

\def\section{\@startsection{section}{1}{\z@}{3.5ex plus 1ex minus
   .2ex}{2.3ex plus .2ex}{\large\bf}}
   \def\thesection{\arabic{section}}

\def\appendix{\setcounter{section}{0}
        \def\thesection{Appendix\ \Alph{section}}
        \def\theequation{\Alph{section}.\arabic{equation}}}

\newcommand{\D}{\Delta}
\newcommand{\beq}{\begin{eqnarray}}
\newcommand{\eeq}{\end{eqnarray}}

\newcommand{\hx}{{\hat{\xi}}}
\newcommand{\al}{{\alpha}}
\newcommand{\be}{{\beta}}

\newcommand{\pa}{{\partial}}

\newcommand{\de}{{\delta}}
\newcommand{\si}{{\sigma}}
\newcommand{\G}{{\Gamma}}
\newcommand{\no}{{\nonumber}}
\newcommand{\f}{\frac}

\begin{document}
\topmargin 0pt
\oddsidemargin -3.5mm
\headheight 0pt
\topskip 0mm
\addtolength{\baselineskip}{0.20\baselineskip}
\begin{flushright}
 CCNY-HEP-01/08 \\
   hep-th/0111224
\end{flushright}
\vspace{0.1cm}
\begin{center}
  {\large \bf  The Hamiltonian Dynamics of Bounded Spacetime and
Black Hole Entropy: The Canonical Method}
\end{center}
\vspace{0.1cm}
\begin{center}
 Mu-In Park\footnote{Electronic address:
muinpark@yahoo.com}
\\{ Department of Physics, City College of the CUNY}\\{ New York, NY 10031}
\end{center}
\vspace{0.1cm}
\begin{center}
  {\bf ABSTRACT}
\end{center}
From first
principles, I present a concrete realization of Carlip's idea on the
black hole entropy from the conformal field theory on the
horizon in any dimension. New formulation
is free of inconsistencies encountered in Carlip's.
By considering
a correct gravity action, whose variational principle is well defined at the
horizon, I $derive$ a correct $classical$ Virasoro generator for the surface
deformations at the horizon through the canonical method.
The existence of the classical Virasoro algebra is
crucial in obtaining an operator Virasoro algebra, through canonical
quantization, which produce the
right central charge and conformal weight $\sim A_+/\hbar G$
for the semiclassical black
hole entropy. The coefficient of proportionality depends on the choice of
ground state, which has to be put in by hand to obtain the correct numerical
factor $1/4$ of the Bekenstein-Hawking (BH) entropy.
The appropriate ground state is different for the rotating and the
non-rotating black holes but otherwise it has a $universality$ for a wide
variety of black holes.
As a byproduct of my
results, I am led to conjecture that {\it non-commutativity of taking
the limit to go to the horizon and computing variation is proportional to
the Hamiltonian and momentum constraints}.
It is shown that almost all the known uncharged black hole
solutions satisfy the conditions for the universal entropy formula.
\vspace{0.1cm}
\begin{flushleft}
9 April 2002 \\
\end{flushleft}
\newpage

\section{Introduction}

In the recent six years, there have been several outstanding approaches
toward a statistical mechanical computation of the
Bekenstein-Hawking (BH) entropy
\cite{Carl:95,Stro:96,Asht,Stro:98}. (See also Ref. \cite{Teit:96} for
an earlier work.) But there is no complete and
consistent understanding about the statistical origin of black
hole entropy so far. Each approach assumes a specific model in a
certain regime but a universal mechanism, which can be applied to any
kind of black holes, had been unclear until the recent seminal work of Carlip
\cite{Carl:99}. According to Carlip, the symmetries of a ``horizon'',
which is treated as a boundary, is a universal mechanism for black hole
entropy. This generalizes Strominger's
approach \cite{Stro:98}, in which the statistical entropy of
the BTZ black hole \cite{BTZ} is computed from the $classical$ Virasoro algebra at
the boundary at ``infinity'' \cite{Brow:86,Bala:99}. By looking at surface
deformations \cite{Teit,Brow:86,Guve} of the ``$r-t$'' plane that
leave the horizon fixed, he has shown that the symmetry algebra contains
a classical Virasoro algebra independently
of spatial dimensions when appropriate boundary conditions are
chosen. With the aid of Cardy's formula for the asymptotic
states, the BH entropy was derived as the leading term of a steepest
descent approximation. The
relevance of the $r-t$ plane at the horizon to black hole
entropy, which resembles the Euclidean gravity formulation in a
radial slicing \cite{Brow:90}, was a key observation: This made it possible
to elevate Strominger's idea to an arbitrary black hole in higher
dimensions without requiring any microscopic model for
quantum states.

There are two big differences between Carlip's approach and
Strominger's, apart from the difference in spacetime dimension. The
first is that
Carlip's Virasoro algebra is computed at the horizon, in contrast to
Strominger's, which is computed at the boundary at infinity where
there is no horizon. Carlip's approach can encode the details of the
metric at the horizon unlike the Strominger's.
The second is that only
``one'' copy of the Virasoro algebra which lives at the
horizon is involved, in contrast to the ``two'' copies of the Virasoro
algebra which live at the boundary at infinity. Similar results have been
subsequently derived in other frameworks as dimensionally reduced
gravity and covariant phase space methods \cite{Solo,Carl:99''}.

Unfortunately it has been known that Carlip's formulation is not
complete \cite{Park:99',Carl:99'''} and there has been no complete
resolution so far \cite{Das:01},
as far as I know, though his idea has received wide interest. This
paper addresses a resolution of the problem.
This provides a concrete realization of Carlip's idea from first
principles.

In section 2, a new canonical Hamiltonian which satisfies the usual
variational
principle in the presence of the boundary is $derived$ from an action
principle for Einstein gravity in any dimension. Full (bulk+boundary)
diffeomorphism ($Diff$) generator is derived immediately from a known
theorem in gravity theory.
Carlip's  Hamiltonian and
$Diff$ generator are also derived from an action principle with a
different choice of boundary action term, but the variational
principle is not well defined in general.

In section 3, a general black hole metric in any dimension is
introduced with the
suitable metric fall-off conditions at the horizon.
Fall-off-preserving conditions are derived for $Diff$
symmetry, which restricts the sub-leading terms as well as
the leading terms for the metric and the $Diff$ parameters.

In section 4, a grand canonical ensemble, in which the horizon temperature
and angular velocity as well as the horizon itself are fixed, is
introduced.  As an immediate consequence of the ensemble and
the fall-off conditions, the new Hamiltonian satisfies the
usual variational principle quite well at the black hole horizon, which is
differentiable in the usual terminology.

In section 5, other consequences of the grand canonical ensemble to the
$Diff$ at the horizon are investigated. First, the new $Diff$
generator satisfies all the conditions for a (differentiable) variational
principle. Second, Carlip's two main assumptions are
$derived$ which leads to a Virasoro-like algebra from the
surface deformation algebra at the horizon: From the condition of fixed
horizon and its temperature, I derive the equation which expresses
radial $Diff$ parameter
$\xi^r$ in terms of the (time or angular) derivatives of temporal or angular
$Diff$ parameter $\xi^t,\xi^{\alpha}$; from the condition of fixed
horizon angular velocity, I derive the equation of zero angular
surface deformations $\hat{\xi}^{\alpha}=0$, which reduces the deformation
algebra to the ``r-t'' plane in the surface deformation space.  Now,
with the help of explicit spacetime dependence on $\xi^t$, which is
inspired by the
null surface at the horizon as well as Carlip's two derived
equations, I show that the surface deformation algebra becomes a Virasoro
algebra with a classical central extension. In deriving this result, I
find, after a tedious computation, a peculiar property that {\it first
  taking the limit} to go
to the horizon and {\it then computing variation} do $not$ commute
by the amount of the Hamiltonian constraint. This leads me to
conjecture that \\

{\it the non-commutativity of the two limiting process
 is proportional to the Hamiltonian and momentum constraints. } \\

By restricting to the case of only one independent rotation, I obtain
the usual momentum space
Virasoro algebra with the classical central charge $c$ and the
conformal weight $\D$ which are proportional to
the horizon area $A_+$.

In section 6, a non-rotating black hole is analyzed with slight
modifications of the formulas for a rotating black hole. The main
difference is the appearance of an additional factor ``1/2'' in the
formula compared to the rotating horizon. There is no
$natural$ connection between the temporal derivative and the angular
derivative of $Diff$ parameters $\xi$.
But one can still introduce an arbitrary velocity parameter, which is
not related
the horizon's rotation, to obtain a Virasoro algebra. Its $c$ and
$\D$  have additional factor ``2'' compared to the
rotating case.

In section 7, the canonical quantization of the Virasoro algebra is
considered. I find that the existence of the {\it classical} Virasoro
algebra is
crucial in obtaining an operator Virasoro algebra which produces the
right central charge and conformal
weight $\sim A_+/\hbar G$ for the semiclassical black hole entropy.
Quantum corrections of operator ordering have negligible
contribution $O(1/\sqrt{\hbar})$ for a large semiclassical BH entropy. In
order to obtain the correct numerical factor $1/4$ of the BH entropy,
the minimum
conformal weight $\D_{\hbox{\scriptsize min}}$, which fixes the ground
state of a black hole configuration, of the classical
Virasoro algebra
 is assumed differently for rotating and non-rotating black holes: For
the rotating one, it is assumed that $\D_{\hbox{\scriptsize
min}}=-(32 \pi G)^{-1} 3A_+ \beta/T$, which has an explicit
dependence on $\beta/T$ with the inverse Hawking temperature
$\beta$ and a temporal-period $T$. For the non-rotating one, a
$\beta, T$ independent ground state  $\D_{\hbox{\scriptsize
min}}=0$ can be assumed. The arbitrary $\beta/T$ dependences in
$c$ and $\D$ exactly cancel each other through {\it effective}
ones $c_{\hbox{\scriptsize eff}}$ and $\D_{\hbox{\scriptsize
eff}}$ such as the final entropy has no $\beta/T$ dependence. The
appropriate ground state is different for the rotating and the
non-rotating black holes but otherwise it has a $universality$ for
a wide variety of black holes.

In section 8, several applications are considered and it is found that
almost all the known uncharged solutions satisfy a universal statistical entropy
formula, which is the same as the BH entropy. The theory is extended
to include the cosmological constant term and it is found that the BTZ
solution and the rotating de-Sitter space have the universal BH entropy
also. This computation is contrary to the other recent computations at the
spatial infinity where there is no horizon and which, therefore, may not be
relevant to a black hole. Especially for the
rotating de-Sitter solution in $n=3$ (Kerr-dS$_3$), no complex number appears at any
intermediate step in contrast to the other computations at spatial infinity,
which is hidden inside the horizon.

I conclude with remarks on several remaining questions which are under
investigation.

I shall adopt unit in which $c=1$.

\begin{section}
{ The variational principle and surface deformations for bounded
  spacetime: A general treatment }
\end{section}

\begin{subsection}
{The variational principle for bounded spacetime in general}
\end{subsection}

Let me start with the spacetime split of the Einstein-Hilbert action
on a $n-$dimensional manifold ${\cal M}$, accompanied by the
extrinsic curvature terms on the boundary $\partial{\cal M}$
\cite{Arno,York:86}
\beq
S=S_{\hbox{\scriptsize EH}} + S_{\partial{\cal M}},
\eeq
where\footnote{$
                 g_{\mu \nu}=\left( \begin{array}{cc}
                 -N^2 +N_aN^a & N_b  \\
                      N_a & h_{ab} \\ \end{array}\right),
                 g^{\mu \nu} =\left(\begin{array}{cc}
                  -N^2 & N_2 N^b \\
                   N^{-2}N^a &h^{ab}-N^{-2}N^a N^b \\ \end{array}\right),
          h_{ac} h^{cb} =\delta_a^b, N^a=h^{ab} N_b $. I follow the
          convention of Wald \cite{Arno} with an exception of (\ref{e}).}
\begin{eqnarray}
\label{SEH}
S_{\hbox{\scriptsize EH}}&=&\frac{1}{16 \pi G}\int_{\cal M} d^n x \left\{ N \sqrt{h}
  \left[ R + \frac{(16 \pi G)^2}{h} \left( p_{ab}
p^{ab} -\frac{1}{n-2} p^2 \right) \right]\nonumber \right. \\
 &&~~~~~~~~~~~~~~~~~\left.+ 2 \partial_t (\sqrt{h}
      K )
+ 2 \partial_a ( - \sqrt{h} K N^a -\sqrt{h} h^{ab}
      \partial_b N ) \right\}, \\
S_{\partial{\cal M}} &=&-\frac{1}{8 \pi G} \int d^{n-1} x ~\sqrt{h} K
  \left|^{\Sigma_{t_f}}_{\Sigma_{t_i}} \right.
+\frac{1}{8 \pi G} \int^{t_f}_{t_i} dt \oint_{\cal B} d^{n-2} x
  ~\sqrt{\sigma} n^a N_a K.
\end{eqnarray}
Here, the boundaries are $\partial{\cal M}=\Sigma_{t_f} \cup
\Sigma_{t_i}\cup {\cal B}$ with the spacelike boundaries
$\Sigma_{t_f}$ and $\Sigma_{t_i}$ at the final and
initial times, respectively and the intersection
${\cal B}$ of an arbitrary timelike boundary with
a time slice ${\Sigma}_t$.\footnote{Here, I assume that the
  spacelike hypersurface $\Sigma_t$ intersects orthogonally the
  timelike boundary ${\cal B}$ for each $t$. See Hawking and Hunter's
  paper \cite{Hawk:96} for a generalization to a non-orthogonal
  intersection.}
$N$ and $N^a ~(a=1,2,\cdots, n-1)$ are the lapse and shift functions,
respectively. $h_{ab}$ is the induced metric
on $\Sigma_t$ and $h$ is its determinant. $R^{ab}$ and $K^{ab}$
are the intrinsic and extrinsic curvature tensors of the
hypersurface $\Sigma_t$, and $R=h_{ab}R^{ab}$ and $K=h_{ab}
K^{ab}$ are their curvature scalars respectively. $p^{ab}=(16 \pi
G)^{-1} \sqrt{h} (K^{ab}-K h^{ab})$ is the canonical momentum
conjugate to $h_{ab}$ and $p=p^{ab} h_{ab}$. $n^a$ is the unit
normal $(h_{ab} n^a n^b =1)$ to the boundary ${\cal B}$ on a
constant time slice $\Sigma_t$, and $\sigma_{ab} =h_{ab}-n_a n_b$
and $\sigma$ are the induced metric on the boundary ${\cal B}$ and
its determinant, respectively. Then the first and the second
boundary terms of $S_{\partial{\cal M}}$ cancel the first and the
second total derivatives terms of $S_{\hbox{\scriptsize EH}}$,
respectively, which are proportional to $K$. The first-order total
derivative action $S$ is closely related to the so-called
``gamma-gamma'' action which eliminates all the second derivatives
of $g_{\mu \nu}~(\mu=0, 1, \cdots, n)$,
but the advantage of the action $S$ is that this can be written in a
manifestly covariant form and moreover the first-order time derivatives of
$N, N^a$ are removed from the start.

For a general hypersurface $\Sigma_t$, the variation of $S$ becomes
\begin{eqnarray}
\label{delS}
\delta S =\frac{1}{16 \pi G} \int_{\cal M} G_{\mu \nu} \delta g^{\mu
  \nu}~ {}^{(n)}{\epsilon} +\frac{1}{16 \pi G} \int _{\partial{\cal
    M}} d \bar{\theta},
\end{eqnarray}
where ${\bar{\theta}}$ is the pull-back of
\begin{eqnarray}
{\theta}=\nabla_{\mu} ({\gamma^{\mu}}_{\nu} \delta r^{\nu} )~
{}^{(n-1)}{\epsilon} \pm \delta r^{\mu} r_{\nu} \nabla^{\nu}
r_{\mu} ~{}^{(n-1)}{\epsilon} + 16 \pi G {\bf P}^{\mu \nu} \delta \gamma_{\mu
  \nu}
\end{eqnarray}
on $\partial {\cal M}$\cite{Burn}.
Here, the unit normal vector $r^{\mu}$ of $\partial {\cal M}$ is
normalized as $g_{\mu \nu} r^{\mu} r^{\nu} =\pm 1$ (the upper sign for a
timelike boundary and the lower sign for a spacelike boundary) and the
induced volume element and the metric on $\partial {\cal M}$ are given
by
\begin{eqnarray}
\label{e}
{}^{(n-1)}\epsilon_{\mu_1 \mu_2 \cdots \mu_{n-1}}=\pm r^{\mu}~
{}^{(n)}\epsilon_{\mu \mu_1 \mu_2 \cdots \mu_{n-1}},
~~\gamma_{\mu \nu} =\mp r_{\mu}r_{ \nu} +g_{\mu \nu}.
\end{eqnarray}
Here, ${\bf P}^{\mu \nu} =-(\Theta ^{\mu \nu}- \Theta \gamma^{\mu
  \nu})~{}^{(n-1)} {\bf\epsilon}$ is the
canonical momentum $(n-1)$-form conjugate to $\gamma_{\mu \nu}$ with the extrinsic
curvature $\Theta^{\mu \nu} =\gamma^{\mu \rho} \nabla_{\rho}
r^{\nu}$. It is clear that, if the induced metric $\gamma_{\mu \nu}$
and the unit
normal $r^{\mu}$ are fixed on $\partial{\cal M}$, the equation of
motion for $g^{\mu \nu}$ is the usual Einstein equation $G_{\mu
  \nu}=0$   at the boundary as well as in the bulk. On the other
hand, since there is
no $physical$ evidence of modification of the Einstein equation at the
boundary in any case, I use the guiding principle,
for the general
treatment of bounded systems, that {\it the equation of motion is the usual
Einstein equation for any kinds of variations}. The boundary
condition, which specifies what quantities are fixed at the boundary,
changes as the boundary term $S_{\partial{\cal M}}$ changes. But it
will be shown that my choice of $S_{\partial{\cal M}}$ is the right one
which is relevant to the black hole horizon when the horizon is treated as
a boundary. However, in this case I do $not$ require the fixed induced
metric or the fixed unit normal at the horizon but rather some appropriate
fall-off conditions for them.
I am going to treat this problem within the Hamiltonian formulation,
which fits my purpose well; in general, the required boundary
condition in the action formulation is different from that of the
Hamiltonian formulation.

The canonical Hamiltonian becomes\footnote{This Hamiltonian form was
  first studied by Brown, Martinez and York \cite{Brow:91} within the
  context of
  thermodynamic partition function. But the physical content is
  different to that of this paper because of different boundary conditions.
See also \cite{Solov} for another literature on the Hamiltonian.}
\cite{Brow:91,Carl:99}
\begin{eqnarray}
\label{H}
H[N, N^a]
&=&\int_{\Sigma}  d^{n-1} x ~(N {\cal H}_t
+N^a {\cal H}_a )
+\frac{1}{8 \pi G}\oint_{\cal B} d^{n-2} x~ ( 16 \pi G N^a {p^{r}}_{a} +
\sqrt{\sigma} n^a D_a N ) \nonumber \\
&\equiv & H_{\Sigma} [N, N^a] +H_{\cal B} [N, N^a],
\end{eqnarray}
where $H_{\Sigma}[N, N^a]$ and $H_{\cal B}[N, N^a]$ are the bulk and
boundary terms on $\Sigma$ and ${\cal B}$, respectively.
${\cal H}_t$ and ${\cal H}_a$ are the ``bulk'' Hamiltonian
and momentum constraints
\begin{eqnarray}
{\cal H}_t =-\f{\sqrt{h}}{16 \pi G} R +\frac{16 \pi G}{\sqrt h} \left(p_{ab} p^{ab}
-\frac{1}{n-2} p^2 \right), ~~{\cal H}_a =-2 D_b {p^{b}}_{a},
\end{eqnarray}
where $D_a$ denotes the covariant derivative with respect to the
 spatial metric $h_{ab}$.
 The variation in $H[N, N^a]$ due to arbitrary variations in
 $h_{ab}, n^a,N, N^a$ becomes
\begin{eqnarray}
\delta H [N, N^a] &=&\delta H_{\Sigma} [N, N^a]+\delta H_{\cal B}
[N,N^a] \nonumber \\
&=&\mbox{bulk ~terms} +\frac{1}{8 \pi G} \int _{\cal B} d^{n-2} x
\left(\delta n^r \partial_r N \sqrt{\sigma} \right. \nonumber \\
\label{delH}
&&\left.+ n^r \partial_r \delta N
  \sqrt{\sigma} + \frac{1}{2} n^r N \sqrt{\sigma} \sigma^{\alpha
    \beta} D_r \delta \sigma_{\alpha \beta}
+16 \pi G \delta N^a {p_a}^r\right)
\eeq
with
\beq
\delta H_{\Sigma} [N, N^a]&=&\mbox{bulk terms} +\frac{1}{8 \pi G} \int
  _{\cal B} d^{n-2} x
\left( \frac{1}{2} n^r N \sqrt{\sigma} \sigma^{\alpha \beta} D_r
  \delta \sigma_{\alpha \beta} \right.\no\\
 &&\left.-\frac{1}{2} n^r D_r N \sqrt{\sigma} \sigma^{\alpha \beta} \delta
  \sigma_{\alpha \beta}
-16 \pi G N^a \delta {p_a}^r \right),\\
\delta H_{\cal B} [N, N^a] &=&\frac{1}{8 \pi G} \int _{\cal B} d^{n-2} x
\left(\delta n^r \partial_r N \sqrt{\sigma} + n^r \partial_r \delta N
  \sqrt{\sigma} + \frac{1}{2} n^r D_r N \sqrt{\sigma} \sigma^{\alpha
  \beta}  \delta \sigma_{\alpha \beta} \right. \\
&&\left.~~~~~~~~~~~~~~~~~~~~~+16 \pi G N^a \delta {p_a}^r  + 16 \pi G \delta N^a
  {p_a}^r \right) , \nonumber
\end{eqnarray}
where I have chosen a coordinate system of
$n^a=(n^r,0,\cdots,0)$\footnote{Greek letters from the beginning of
  the alphabet are the boundary indices which do not include radial
  coordinate $r$ whence Greek letters from the middle of the alphabet
  are spacetime indices \cite{Carl:99}. }.

The  bulk terms are the usual variation terms for $\delta p^{ab},
\delta h_{ab}$, which produce the bulk equations of motion
\cite{Arno} as well as the Hamiltonian and momentum constraints
\begin{eqnarray}
\label{Ht=0}
{\cal H}_t \approx 0, ~~
{\cal H}_a \approx 0.
\end{eqnarray}

The additional variations at the boundary could affect the bulk equation of
 motions in general. Let me first consider the first
 term in the boundary terms of (\ref{delH})
. By using, from the definition of $n^c$,
\begin{eqnarray}
\label{deln}
\delta n^c =-\frac{1}{2} \delta h_{ab} h^{cb} n^a
\end{eqnarray}
the first term becomes
\begin{eqnarray}
-\frac{1}{16 \pi G} \int _{\cal B} d^{n-2} x~ \delta h_{ab} ( h^{rb}
 n^a \partial_r N \sqrt{ \sigma} ).
\end{eqnarray}
So, this term would produce an additional term in the bulk equation of motion
of the form
\begin{eqnarray}
\label{{pi,H}}
\{p^{ab} (x), H[N, N^a] \} |_{\hbox{\scriptsize boundary}} =\frac{1}{16 \pi G}
\delta (r-r_{\cal B}) (n^a h^{br} \partial_r N \sqrt{\sigma} )
\end{eqnarray}
unless the quantity in the bracket (~) vanishes on the boundary ${\cal B}$.
Here $r_{\cal B}$ is the radius of the boundary ${\cal B}$.
Similarly, the last boundary term in (\ref{delH}) produces an additional
contribution
\begin{eqnarray}
\label{Habound}
{\cal H}_a |_{\hbox{\scriptsize boundary}} =\delta (r-r_{\cal B})2 {p^r}_a
\end{eqnarray}
to the momentum constraints ${\cal H}_a$ in general. Contrary to
these two contributions, which are proportional to $\delta
(r-r_{\cal B})$, the second and third terms produce a highly
singular term $\partial_r \delta (r-r_{\cal B})$, which produces
the divergence of ``$\delta(0)$'' even when I compute the
commutation relations between the integrated quantities $\{H[N,
N^a], H[N', N^{a'}] \}$, which should be related to the measurable
things. In order to avoid this problem, I need to assume the
boundary conditions which restrict the radial derivatives of the
variations $\delta N$ and $\delta \sigma_{\alpha \beta}$ on the
boundary ${\cal B}$ as
\begin{eqnarray}
\label{Nbound}
&&\sqrt{\sigma} n^r \partial_r \delta N |_{\cal B} =0, \\
\label{sbound}
&&N \sqrt{\si} \sigma^{\alpha \beta} n^r D_r \delta \sigma_{\alpha \beta}
\left|_{\cal B}\right. =0
\end{eqnarray}
but $N$ and $\sigma_{\alpha \beta}$ can be arbitrary otherwise.
One may also take, instead of (\ref{Nbound}, \ref{sbound}),
$\delta N|_{\cal B} =0$ and $\delta \sigma_{\alpha \beta}|_{\cal
B} =0$ but this could be too strong a condition depending on the
property of ${\cal B}$. Hence I keep (\ref{Nbound}) and
(\ref{sbound}) for a general treatment.

Now, in summary, about the contributions of the boundary terms in
(\ref{delH}), there are two possible contributions (\ref{{pi,H}}) and
(\ref{Habound}) to the corresponding bulk ones from the first and last
terms in (\ref{delH}), respectively. The two badly-behaving terms are
removed by
restricting $\partial_{r} \delta N$ and $\partial_r \delta \sigma_{\alpha
  \beta}$ as in (\ref{Nbound}) and (\ref{sbound}).
In other words, the Hamiltonian (\ref{H}) produces the Einstein
equation on the boundary ${\cal B}$ ``as well as '' in the bulk
 when the contributions of (\ref{{pi,H}}) and (\ref{Habound}) vanish, and
 the boundary conditions
 (\ref{Nbound}) and (\ref{sbound}) are satisfied on ${\cal B}$ ($g_{ab}$ is
fixed on the boundaries at infinity $\Sigma_{t_f, t_i}$ as usual such that
there is no contributions from $\Sigma_{t_f, t_i}$).

For completeness, let me remark on another interesting type of boundary
${\cal C}$ where there is $no$ added boundary action
$S_{\partial{\cal M}}$
such as when the total action is just the Einstein-Hilbert action
$
S=S_{EH}.
$
In this case, ${\bf \theta}$ in (\ref{delS}) becomes
\begin{eqnarray}
\label{thetaC}
{\bf \theta} |_{\cal C} =&&- \frac{2}{n-2}16 \pi G \gamma_{\mu \nu} \delta
{\bf P}^{\mu \nu} + (1 -\frac{2}{n-2} )16 \pi G {\bf P}^{\mu \nu} \delta \gamma_{\mu
  \nu} \nonumber \\
&&+ \nabla_{\mu} (
{\gamma^{\mu}}_{\nu} \delta r^{\nu} )~ {}^{(n-1)} {\bf \epsilon} \pm
\delta r^{\mu} r_{\nu} \nabla^{\nu} r_{\mu}~ {}^{(n-1)} \epsilon .
\end{eqnarray}
This, in general, contains $\delta {\bf P} ^{\mu \nu}$ term as
well as $\delta \gamma_{\mu \nu}$ and $\delta r^{\mu}$ terms.
Hence, for the boundary ${\cal C}$, one must fix ${\bf P}^{\mu
\nu}$ as well as $\gamma_{\mu \nu}$ and $r^{\mu}$ on it, which
may in turn over-specify the boundary degrees of freedom. The
canonical Hamiltonian becomes, for a timelike boundary ${\cal C}$,
\begin{eqnarray}
H'[N, N^a] &=& \int_{\Sigma}  d^{n-1} x ~(N {\cal
  H}_t +N^a {\cal H}_a )  \nonumber \\
\label{H'}
&&~~+\frac{1}{8 \pi G}\oint_{\cal C} d^{n-2} x
\left(\frac{16 \pi G}{2-n}\sqrt{\frac{\sigma}{h}} n^a N_a p+ 16 \pi G
  N^a {p^{r}}_{a} +
\sqrt{\sigma} n^a D_a N \right) .
\end{eqnarray}
This has the same form that Carlip has
``postulated'', which can be seen by expressing the first $p$ term
in terms of
extrinsic curvature
$K=16 \pi G ((2-n) \sqrt{h})^{-1} p$ \cite{Carl:99}.
But it has a drawback that the variational principle is not well defined
in general (see Appendix
{\bf A} for details; see also Ref. \cite{Park:99'}), which would be
related to the over-specification of boundary data in (\ref{thetaC}). So,
I will not consider this boundary ${\cal C}$ anymore hereafter and
concentrate only the boundary ${\cal B}$.\\

\begin{subsection}
{$Diff$ generators}
\end{subsection}

Now, let me consider the generators of the spacetime
$Diff$
\begin{eqnarray}
\label{diff}
 \delta_{\xi} x^{\mu}=-\xi^{\mu},
 ~~\delta_{\xi} g_{\mu \nu}=\xi^{\sigma} \partial_{\sigma} g_{\mu \nu} +
\partial_{\mu} \xi^{\sigma} g_{\sigma \nu} +\partial_{\nu}
\xi^{\sigma} g_{\sigma \mu}.
\end{eqnarray}
The generators may be obtained directly from the usual Noether
procedure \cite{Jack,Oh:98,Park:99,Oh:99}. But there is a well-known
theorem which identifies the generators from a well-defined
Hamiltonian \cite{Guve}:
If $H[N, N^a]$ is a canonical Hamiltonian of the gravity theory
which does not have boundary
term in the variations $\delta H[N, N^a]$ with
the lapse and shift functions $N$ and $N^a$, the $Diff$ generator
$L[\hat{\xi}]$ of
\begin{eqnarray}
\label{delhp}
\delta_{\xi} h_{ab} =\{ h_{ab}, L[ \hat{\xi} ] \},
~~\delta_{\xi} p^{ab} =\{ p^{ab}, L[\hat{ \xi}] \}
\end{eqnarray}
is given by substituting $N, N^a$ in the Hamiltonian $H[N, N^a]$ with
the so-called surface deformation parameters \cite{Teit,Brow:86,Guve}
\begin{eqnarray}
\label{xhat}
\hat{\xi}^t =N \xi^t, ~\hat{\xi}^a =\xi^a +\xi^t N^a
\end{eqnarray}
respectively, giving
\begin{eqnarray}
L[\hat{\xi}]=H[N, N^a]_{(N, N^a)\rightarrow (\hat{\xi}^t, \hat{\xi}^a)
  }.
\end{eqnarray}
 The $Diff$ of $\delta_{\xi} g_{\mu t}$ is instead given by the basic
 formula (\ref{diff}) essentially due to the absence of the canonical
 momentum conjugate to $g_{\mu t}$. Hence the generator becomes (I
 denote $L[\hat{\xi}]=H[\hat{\xi}]$)
\begin{eqnarray}
\label{Hhat}
H[\hat{ \xi}]&=& \int_{\Sigma} d^{n-1} x  ~{\hat
  \xi}^{\mu} {\cal H}_{\mu}
  +J [{\hat \xi}],
\end{eqnarray}
where
\begin{eqnarray}
\label{Jhat}
J [\hat{ \xi}]&=&\frac{1}{8 \pi G} \oint_{{\cal B}} d^{n-2} x~ (
  16 \pi G \hat{\xi}^a {p^{r}}_{a} + \sqrt{\sigma} n^a D_a
  \hat{\xi}^t )
\end{eqnarray}
for the boundary ${\cal B}$, I assume the boundary conditions for
  the well-defined variations $\delta H[N, N^a]$
  without boundary terms\footnote{Of course, this theorem will be
 modified if I allow the boundary terms in the variation
  principle \cite{Tlee}.}.
The usual $Diff$ at the boundary
as well as in the bulk is generated by the cancellation of
the boundary $Diff$ from the bulk part in
(\ref{Hhat}) and another boundary $Diff$ from the boundary
  generator $J[\hat{\xi}]$. Another interesting effect, which is
  crucial to my analysis of black hole
entropy, of $J[\hat{\xi}]$ is that it may produce the
``classical'' central extension in the symmetry algebra in
general \cite{Regg,Brow:86}.

Since I am interested in the symmetry algebra of $H[\hat{\xi}]$
on the $physical~ subspace$ where the Hamiltonian and momentum
constraints of (\ref{Ht=0}) are imposed, I must compute
the Dirac bracket in general \cite{Park:99}. The Dirac bracket algebra
would show some interesting
effects of the boundary but this is very complicated in my
case. Rather, in this paper, I will use an {\it effective} method
which gives the Dirac bracket of $H [\hat{\xi}]$'s themselves without
tedious computations \cite{Regg,Brow:86}\footnote{This can be
  equivalently written as $\{ J[\hat{\xi}], J[\hat{\eta}]\}^*
  =(\delta_{\eta} J[\hat{\xi}]-\delta_{\xi} J[\hat{\eta}] )/2$ in
  order that the antisymmetry under $\eta \leftrightarrow \xi$ is
  manifest.  This can be realized through an explicit form of the
  Dirac bracket also.}
\begin{eqnarray}
\label{RT}
\{H[\hat{\xi}], H[\hat{\eta}] \}^* &\approx& \{J[\hat{\xi}],
J[\hat{\eta}] \}^* =\delta_{\eta} J [\hat{\xi}]
\end{eqnarray}
from the definition of the Dirac bracket. Here, the Dirac bracket
on the right hand side has the form
\begin{eqnarray}
\label{{J,J}}
\{J[\hat{\xi}], J[\hat{\eta}] \}^*=J[\{ \hat{\xi}, \hat{\eta} \}_{\hbox{\scriptsize SD}} ]
+ K[ \hat{\xi}, \hat{\eta}]
\end{eqnarray}
in general, where $K[\hat{\xi}, \hat{\eta} ]$ is a possible central
term and $\{\hat{\xi}, \hat{\eta} \}_{\hbox{\scriptsize SD}}$ is the Lie bracket for the
algebra of surface deformations \cite{Teit,Brow:86,Guve}:
\begin{eqnarray}
\{\hat{\xi}, \hat{\eta} \}^t_{\hbox{\scriptsize SD}} =\hat{\xi}^a \partial_a
\hat{\eta}^t-\hat{\eta}^a \partial_a \hat{\xi}^t ,
~~\{\hat{\xi}, \hat{\eta} \}^a_{\hbox{\scriptsize SD}} =\hat{\xi}^b \partial_b
\hat{\eta}^a-\hat{\eta}^b \partial_b \hat{\xi}^a +
g^{ab} (\hat{\xi}^t \partial_b
\hat{\eta}^t-\hat{\eta}^t \partial_b \hat{\xi}^t).
\end{eqnarray}

\begin{section}
{ Fall-off conditions at a black hole horizon }
\end{section}

Now, let me consider a general black-hole-like metric in $n$ spacetime
dimensions \cite{Myer} with the Boyer-Lindquist coordinates $(t, r,
x^{\alpha})$
\begin{eqnarray}
\label{ds2}
ds^2=-N^2 dt^2 +f^2 (dr +N^r dt )^2 +\sigma_{\alpha \beta} (d
x^{\alpha} +N^{\alpha} dt) (d x^{\beta} + N^{\beta} dt),
\end{eqnarray}
where the lapse function $N$ vanishes at the horizon and behaves
as follows near the ``outer-most'' horizon $r=r_+$
\begin{eqnarray}
N^2=h(x^{\alpha}) (r-r_+) + O (r -r_+)^2,~~\label{beta}
\frac{2 \pi}{\beta}=n^a \partial_a N|_{r_+},
\end{eqnarray}
where $\beta$ is the inverse Hawking temperature, which is
constant on the horizon $r_+$, and $n^a$ is the unit normal to the
horizon boundary $r=r_+$ on a constant time slice $\Sigma_t$.

The suitable fall-off conditions near the horizon $r_+$ are
\begin{eqnarray}
\label{f=N-1}
&&f=\frac{\beta h}{4 \pi} N^{-1} +O (1), \\
\label{Nr}
&&N^r=O (N^{2}),~~
(\partial_t -N^r \partial_r )g_{\mu \nu} =O (N) g_{\mu \nu}, \\
\label{sigma}
&&\sigma_{\alpha \beta}=O (1), ~~
N^{\alpha} =O (1), \\
\label{DN}
&&D_{\alpha} N_{\beta} +D_{\beta}N_{\alpha}=O(N).
\end{eqnarray}
The conditions in (\ref{Nr}) are the perturbations of the
stationary black hole of $N^r=0$; I shall keep $N^r$ in this and
the next sections for generality; but $N^r=0$ shall be considered
in section 5.c by computing the Virasoro algebra for a stationary
horizon. (\ref{DN}) is the condition of constant angular velocity
$\Omega^{\alpha}=-  N^{\alpha}$. However by comparing with
(\ref{sigma}) one can see that (\ref{DN}) gives a nontrivial
restriction if one considers $D_{\alpha} \sim \partial
_{\alpha}\sim O(1)$. To make it more explicit, let me consider a
decomposition
\begin{eqnarray}
\label{N=f+K}
N_{\alpha}=f_{\alpha} +K_{\alpha},
\end{eqnarray}
where
$ f_{\alpha} =O (1),~ D_{\alpha} f_{\beta} +D_{\beta} f_{\alpha} =0$,
and $K_{\alpha}$ vanishes at the horizon $r_+$. Then, it is easy
to see that (\ref{DN}) is satisfied if
\begin{eqnarray}
\label{drsigma}
\partial_r \sigma_{\alpha \beta} \leq O(N^{-1}), ~~
K^{\alpha} \leq O(N).
\end{eqnarray}
[Note that $\partial_r N =2 \pi f {\beta}^{-1}$.] Therefore, I
require the form (\ref{N=f+K}), with the condition
(\ref{drsigma}), for consistency with (\ref{f=N-1})-(\ref{DN}).
These are the basic set-up of the fall-off conditions at the
horizon $r_+$.

As another requirement of consistency, the $Diff$ symmetry
(\ref{diff}) should not accidently violate these fall-off
conditions. The suitable conditions for this requirement
are\footnote{Notice that the hat of
  $\hat{\xi}^{\alpha}$ in (\ref{drxt}) is not mistyping. In the
  unhatted expression, it becomes $\partial_r \xi^{\alpha}
  =-N^{\alpha} \partial_r \xi^t + O(1)$, which is important when I
  prove the condition (\ref{sbound}) in $Diff$ (\ref{diffsbound}).}
\begin{eqnarray}
\label{xt}
&&\xi^t \leq O (1), ~~
\xi^{\alpha} \leq O (1), ~~
 \xi^r \leq O (N^2), \\
\label{drxt}
&&\partial_r \xi^t \leq O (N^{-2}), ~~
\partial_r \hat{\xi}^{\alpha} \leq O(1), \\
\label{drNbeta}
&&\partial_r N^{\beta} \leq O (1),~~
\partial_r \sigma_{\alpha \beta} \leq O (N^{-2}).
\end{eqnarray}
This shows that the fall-off-condition-preserving $Diff$ requires the
suitable behaviors of the sub-leading terms of $Diff$ parameters
from (\ref{drxt}) as well as the behavior of the
leading terms (\ref{xt}). Moreover,
it depends also upon the behavior of the sub-leading term of the metric
from (\ref{drNbeta}).
Then, it is
straightforward to compute that
\begin{eqnarray}
\label{Krr}
K_{rr}=O (N^{-3}) ,~~
K_{\alpha r} =O(N^{-1}), ~~
K_{\alpha \beta}=O (1), ~~
K=O (N^{-1})
\end{eqnarray}
and
\begin{eqnarray}
\label{pirr}
{p_r}^r=O (N^{-1}), ~~
{p_{\alpha}}^r =O (1),~~ {p_r}^{\alpha} =O (N^{-2}), ~~
{p_{\alpha}}^{\beta} =O (N^{-2}), ~~
p= O(N^{-2}).
\end{eqnarray}
Note that there is no inequality relation in contrast to
the fall-off preserving conditions
(\ref{drsigma})-(\ref{drNbeta}).\\

\begin{section}
{ Grand canonical ensemble and the variational principle at the black
  hole horizon}
\end{section}

I have considered the variational principle for a bounded spacetime
in general in the two preceding sections.
Now, let me consider the variational principle at the black hole
horizon $r_{+}$ when the horizon is treated as a boundary with the
fall-off boundary conditions as described in the last section.
But, for discussing the statistical mechanical property of a black hole, I
must specify the type of ensemble first of all. For this purpose, I take
a $grand~canonical$ ensemble, in which {\it the horizon $r_+$ and
  its temperature and angular
velocity are fixed} \cite{Brow:91}. Full
elaboration on the impact of this ensemble will be given in the next
section, but here I only mention a direct consequence for the variational
principle: When the horizon $r_+$ is treated as the boundary ${\cal
  B}$, the boundary condition (\ref{Nbound}) is automatic since
\begin{eqnarray}
\label{delN=0}
\delta N |_{r_{+}}=0
\end{eqnarray}
 in order that the boundary remains at the horizon $r=r_+$, which is
 the solution of $N+\delta N =0$, and also in order that the temperature
 $\beta$, which is the coefficient of $N+\delta N$, is
 unchanged. However, the condition (\ref{sbound}) is not automatic and
 I must impose this for consistency.

Moreover, the fixed angular velocity $\Omega^{\alpha} \sim
N^{\alpha}$ in the ensemble insures that there is no boundary
contributions to the momentum constraint ${\cal H}_{\alpha}$ (\ref{Habound}),
which is the coefficient of $\delta N^{\alpha}$.

One can then see that,
 using the fall-off conditions in (\ref{pirr}), the only
 non-vanishing term in
(\ref{Habound})
is
$
{\cal H}_r |_{\hbox{\scriptsize boundary}}=O(N^{-1}).
$
But this is not harmful because its contribution to the bulk Hamiltonian vanishes:
$
\int _{\Sigma} d^{n-1} x~ N^r {\cal H}_r \leq O(N).
$
On the other hand, the boundary contributions of (\ref{{pi,H}})
vanish also since
\begin{eqnarray}
\{p^{rr}(x), H[N, N^a] \}|_{\hbox{\scriptsize boundary}}
                  \leq O(N^2),
\end{eqnarray}
and all other components in (\ref{{pi,H}}) vanish trivially from my
choice of coordinates and the metric (\ref{ds2}).
Therefore, the new Hamiltonian
(\ref{H}) admits the variational principle even at the black hole horizon
such that the usual $bulk$ equations of motion
and the bulk constraints (\ref{Ht=0}) can be applied to the horizon as
well as outside of the
horizon.

Moreover, I note that since the first term in the boundary part of the
Hamiltonian (\ref{H}) is
$
N^a {p^r}_a =N^{\alpha} {p^r}_{\alpha} +O(N) \leq O(1),
$
the Hamiltonian at the black hole horizon, which will be the true
Hamiltonian on the physical subspace, reduces to
\begin{eqnarray}
\label{Honr+}
H|_{r_+}[N, N^a]&=&\frac{1}{8 \pi G} \oint _{r_+} d^{n-2} x~
(16 \pi G N^{\alpha} {p^r}_{\alpha} +\sqrt{\sigma} n^r D_r N)
\leq O(1).
\end{eqnarray}

\begin{section}
{ $Diff$ and Virasoro algebra at the horizon}
\end{section}

In section 2.b, the general $Diff$ generators and their symmetry algebras have
been considered for an arbitrary boundary ${\cal B}$. But, now
let me restrict to the horizon in particular, where
the general $Diff$ reduces to (\ref{xt})-(\ref{drNbeta}). But, to
this end, I need
more detailed knowledge on the $Diff$ in the grand canonical
ensemble for the black hole. \\

\subsection{$Diff$ at the horizon: Boundary conditions}

First, I note
that the boundary condition (\ref{sbound}) is automatic
\begin{eqnarray}
\label{diffsbound} N \sqrt{\si} \sigma^{\alpha \beta} n^r D_r
\delta_{\xi} \sigma_{\alpha \beta} \leq O(N)
\end{eqnarray}
by using
the condition
(\ref{drsigma}) and the fact that
\begin{eqnarray}
\label{sigma(sol)}
\delta_{\xi} \sigma_{\alpha \beta} (x^{\mu}) = O(N).
\end{eqnarray}

Moreover, the general requirement (\ref{delN=0}) of the fixed horizon and
temperature for the grand canonical ensemble in the variational
principle at the black hole horizon should be applied to $Diff$
transformation also. Now let me elaborate in some details what the further
implications of this for $Diff$. From the fact that $g^{tt} =-1/N^2$ and $\delta_{\xi}
g^{tt} =2 \nabla ^t \xi^t $, one finds
\begin{eqnarray}
\delta_{\xi}N^2
=\frac{4 Nf}{\beta} (\xi^t N^r +\xi^r)+2 N^2 (\partial_t -N^a
\partial_a ) \xi ^t +\xi^{\alpha} \partial_{\alpha} N^2+O(N^3)
\le O(N^2).
\end{eqnarray}
Hence, an arbitrary $Diff$ transformation produces change in the
inverse temperature\footnote{Notice that another term
  $\delta_{\xi} n^r \partial_r N$ from (\ref{beta}) should be dropped
  since this
  gives the temperature at the lifted position $r_{+}(n_r
  +\delta_{\xi} n^r)$ not the horizon $r_+ n_r$.} $\beta^{-1}$ by
$(2 \pi)^{-1} n^r \partial_r
\delta_{\xi}N |_{r_+}=(4 \pi f N)^{-1}\partial_r\delta_{\xi}
N^2|_{r_+}$ unless the condition (\ref{delN=0}) is satisfied for the
$Diff$ transformation also, i.e.,
\begin{eqnarray}
\label{delN+=0}
\delta_{\xi} N^2 |_{r_+} =0.
\end{eqnarray}
In other words, the grand canonical ensemble is satisfied if
\begin{eqnarray}
\label{xr(sol)}
\xi^r=-\frac{N \beta}{2\pi f} (\partial_t -N^{\alpha}
\partial_{\alpha} )\xi^t + \frac{N \beta N^r}{2 \pi f} \left(\partial_r
\xi^t -\frac{2 \pi f}{N \beta} \xi^t \right)-\frac{N^3 \beta}{4 \pi f}
\xi^{\alpha} \partial_{\alpha} g^{tt},
\end{eqnarray}
which allows to express $\xi^r$ in terms of $\xi^t$.\footnote{Notice that
here, there is no perturbation on
  the black hole radius $r_+$ since $\delta_{\xi} N^2 =O(N^2) \sim
  (r-r_+)$; this can be more directly seen from $\delta_{\xi} x^r
  |_{r_+} =-\xi^r |_{r_+}=0$ due to (\ref{diff}) and (\ref{xt}).
 But this
  does $not$ imply the fixed horizon area $A_+=\oint_{r_+} d^{n-2}
  x~\sqrt{\sigma}$, i.e., fixed entropy, which might cause the
  ensemble to be
  absurd: Rather, one has $\delta_{\xi}A_+=\oint_{r_+} d^{n-2}
  x~\delta_{\xi}\sqrt{\sigma}=O(1)$ according to the solution
  (\ref{sigma(sol)} ).}
Hence, another boundary condition (\ref{Nbound}) is also automatic in the
grand canonical ensemble.

Therefore, the $Diff$ generator (\ref{Hhat}), for the
boundary ${\cal B}$ as the horizon, is well-defined without the
additional contributions from the boundary when (\ref{xr(sol)}) as
well as all the other conditions (\ref{Nbound}) and
(\ref{sbound}) for the metric and the conditions for the
$Diff$ parameters
 are satisfied (all the other boundary
contributions vanish as the Hamiltonian did (section 4)). The existence of the
well-defined $Diff$ generator implies,
according to the Noether theorem \cite{Oh:98,Park:99,Oh:99}, that
{\it the
$Diff$ symmetry is not broken even with the boundary}.

Furthermore, with the well-defined generator $H[\hat{\xi}]$
(\ref{Hhat}) all the
information on the $Diff$ of the bulk part $\int_{\Sigma} d^{n-1}x~
\hat{\xi}^{\mu} {\cal H} _{\mu}$ is encoded into the boundary (horizon
$r_+$) part $J [\hat{\xi}]$ when it is treated on the physical
subspace of ${\cal H}_{\mu} \approx 0$; this is another aspect of the
't Hooft {\it holography principle} at the black hole \cite{Thoo}.

\subsection{$Diff$ at the horizon: No angular surface deformations}

In the last subsection I have used the condition of $fixed~
temperature$ for a grand canonical ensemble. Now, let me consider
the condition of $fixed~ angular~velocity$, which is due to a
chemical potential for the black hole system. Since
$\Omega^{\alpha}\sim N^{\alpha}$, I must compute $\delta_{\xi}
N^{\alpha}$ to explore the rotational property of a grand
canonical ensemble. On the other hand, since $
g_{t \alpha} =\sigma_{\alpha \beta} N^{\beta},
$
I can compute $\delta_{\xi} N^\al$ (from
$
\sigma_{\alpha \beta} \delta_{\xi} N^{\beta} =\delta_{\xi} g_{t
  \alpha} -\delta_{\xi} \sigma_{\alpha \beta} N^{\beta})
$
as
\begin{eqnarray}
\label{delNalpha}
\delta_{\xi} N^{\alpha} =\partial_t \hat{\xi}^{\alpha} -N^{\beta}
\partial_{\beta} \hat{\xi}^{\alpha} +\xi^{\delta} \partial_{\delta}
N^{\alpha} +\xi^t N^{\beta} \partial_{\beta} N^{\alpha} + O(N).
\end{eqnarray}
Then, one finds that the condition of fixed angular velocity
$\Omega^{\alpha} \sim N^{\alpha}$, i.e., $\delta_{\xi}
N^{\alpha}|_{r_+}=0$ of a grand canonical ensemble is satisfied
when I restrict the surface deformation space of $\hat{\xi}^{\mu}$
(\ref{xhat}) to the ``r-t'' plane:
\begin{eqnarray}
\label{xalphahat=0} \hat{\xi}^{\alpha}=0;
\end{eqnarray}
 of course,
this does not mean the ``r-t'' plane in the space of {\it spacetime} $Diff$
 $\xi^{\mu}$. Furthermore, the solution is
$unique$ in the grand canonical ensemble\footnote{ The question of
  angular $Diff$ with $\hat{\xi}^{\alpha}\neq 0$ has been raised by
  Carlip in his talk \cite{Carl:99}. I can rule out this in a grand
  canonical ensemble. But it is unclear whether this can be ruled out in
  other ensembles also. }.

Now, equation (\ref{xalphahat=0}), together with the equation (\ref{xr(sol)}),
has been fixed from the conditions of the grand canonical ensemble on the
black hole; these were the two main assumptions in the Carlip's formulation.
So, is this the end of story of the grand canonical ensemble ? To
answer this, let me elaborate what the condition (\ref{xalphahat=0})
further implies. From the definition (\ref{diff}), equation
(\ref{xalphahat=0}), i.e.,
$\xi^{\alpha}=-N^{\alpha} \xi^t$ implies
\begin{eqnarray}
\label{delx}
\delta_{\xi} x^{\alpha}=-N^{\alpha} \delta_{\xi}t
\end{eqnarray}
and so,
\begin{eqnarray}
\label{delx-1}
\Omega^{\hat{\alpha}} \frac{\delta}{\delta_{\xi}
  x^{\hat{\alpha}}}=\frac{\delta}{\delta_{\xi} t} ~~\mbox{(no~ sum)}
\end{eqnarray}
with $\Omega^{\hat{\alpha}} =-N^{\hat{\alpha}}$. This does not show
any definite information about arbitrary spacetime variations. But,
let me introduce one assumption about $\xi^t$ inspired by these
equations (\ref{delx}) and (\ref{delx-1}) that \\

``$\xi^t$ is at rest on the horizon''. \\

This assumption determines the spacetime dependence of
$\xi^t$ as\footnote{For $N^r=0$, a more generalized form $
\xi^t=\xi^t (t-\mu^{-1} r_* +\sum_{\hat{\alpha}}
(\nu^{\hat{\alpha}} \Omega^{\hat{\alpha}})^{-1} x^{\hat{\alpha}},
  x^{\bar{\alpha}} )$ may considered by relaxing the assumption that
  ``$\xi^t$ is at rest on the horizon''. But one can show that taking
  different $\mu, \nu$, and $T/\beta$ corresponds to taking different
  ground state.
}
\begin{eqnarray}
\label{xtassump}
\xi^t=\xi^t \left[t-\left(1-\f{f}{N} N^r \right)r_* +\sum_{\hat{\alpha}}
\frac{1}{\Omega^{\hat{\alpha}}} x^{\hat{\alpha}}, x^{\bar{\alpha}} \right],
\end{eqnarray}
where `$t-(1-f N^{-1} N^r) r_* $=constant' is a radial, outgoing null
geodesic for a generic
Kerr black hole with the Regge-Wheeler tortoise coordinate
\begin{eqnarray}
r_*= (\beta/4 \pi) \mbox{log} (r-r_+) +O(r-r_+)
\end{eqnarray}
near the horizon $r_+$; an appropriate integration constant
for $dr/dr_*=N/f$ is chosen here. The angular dependence on
$x^{\hat{\al}}$ reflects the assumption that $\xi^t$ is ``at rest''
on the horizon which is rotating with velocity $\Omega^{\hat{\al}}$;
however, notice that
$x^{\bar{\alpha}}$ dependence in $\xi^t$ is arbitrary in
general. Now then, one can show that
\begin{eqnarray}
\label{drxt(sol)}
&&\partial_r \xi^t =-\frac{f}{N}\left(1-\f{f}{N} N^r \right)
 \partial_t \xi^t, \\
\label{dtxt}
&&\partial_t \xi^t=\Omega^{\hat{\alpha}} \partial_{\hat{\alpha}}
\xi^t ~~\mbox{(no~sum)},
\end{eqnarray}
which makes it possible to express all the derivatives in terms of one
dimensional derivatives $\partial_t$ or $\partial_{\hat{\alpha}}$.
Here, (\ref{dtxt}) shows that
$\xi^t$ respects the symmetry (\ref{delx}) exactly. Moreover,
(\ref{drxt(sol)}) is consistent with the boundary condition
(\ref{drxt}). One can also show that another non-vanishing $Diff$
parameter $\xi^{\alpha}$ satisfies the same equations as
(\ref{drxt(sol)}) and (\ref{dtxt}).

\subsection{The Virasoro algebra at a stationary horizon}

So far my computation was valid for any $N^r$, which is $O(N^2)$. But,
now let me
focus on a time slice with
$ N^r =0$ such that a ``stationary'' black hole is
considered. Then, the
$Diff$ of $J[\hat{\xi}]$ with respect to the surface deformations
becomes
\begin{eqnarray}
\label{delJ}
\delta_{\xi_2} J [\hat{\xi_1}]&=&\frac{1}{8 \pi G}  \oint_{r_+} d^{n-2} x
\left[ \delta_{\xi_2} n^r \partial_r
\hat{\xi}_{1t} \sqrt{\sigma}
+n^r \partial_r (\delta_{\xi_2} \hat{\xi}^t_1 ) \sqrt{\sigma}
+ \frac{1}{2} n^r \partial_r \hat{\xi}^t_1 \sqrt{\sigma} \sigma^{\alpha \beta}
 \delta_{\xi_2} \sigma_{\alpha \beta} \right. \nonumber  \\
&&~~~\left.+16 \pi G \delta_{\xi_2}\hat{\xi_1}^a p^r_a
+16 \pi G \hat{\xi_1}^a \delta_{\xi_2} p^r_a \right].
\end{eqnarray}
It is straightforward to check that the second to the fourth terms vanish
based on the relations for the metric and the $Diff$ parameters.
The only non-vanishing contributions are the fifth and the first terms.

{\it The fifth term}: This term needs a tedious computation of
$\{p^{ab}, H [\hat{\xi}]\}$ from (\ref{delhp}). But the final
result is very simple (the derivation is sketched in Appendix {\bf
B}): This becomes, from $\hat{\xi}^{\alpha}=0$,
\begin{eqnarray}
\label{fifth}
2 \oint _{r_+} d^{n-2}x~ \hat{\xi_1}^r \delta_2
{p^r}_r =\oint _{r_+} d^{n-2}x~
\hat{\xi_1}^r \hat{\xi_2}^t {\cal H}^t +O(N),
\end{eqnarray}
where
\begin{eqnarray}
{\cal H}^t =\frac{16 \pi G}{\sqrt{h}} \left(p_{\alpha \beta} p^{\alpha \beta}
-\frac{1}{n-2} {p^{\alpha}}_{\alpha} {p^{\beta}}_{\beta}\right)
+O(N^{-2})
\end{eqnarray}
is the Hamiltonian constraint evaluated near the horizon $r_+$. This
result shows a very peculiar property since the $Diff$ of
$\delta_{\xi} {p^r}_r$ does not preserve the fall-off
behavior ${p^r}_r=O(N^{-1})$ near the horizon, but rather becomes
as $\delta_{\xi} {p^r}_r=O(N^{-2})$ such that the left hand side of
(\ref{fifth}) is $O(1)$, which does not vanish in general. Notice that
this is sharply contrary to the process of {\it first taking the
  limit} to go to the horizon in the corresponding term
  $ 2 \oint _{r_+} d^{n-2} {\xi_1}^r {p^r}_r=O(N)=0$ in
$J[\hat{\xi}]$ and {\it then computing the functional
  differentiation }$\delta_{\eta} 0=0$; these two processes do not
commute in general\footnote{ This has been first pointed out by
  Carlip in the interpretation of a result of Ho and I
  \cite{Park:99',Carl:99'''}
.}. However, the result
(\ref{fifth}) shows
the interesting property that the surviving terms are nothing but the
constraints
of the system. Hence, the problematic situation of non-commutativity
of the two limiting processes of computing the variations
and the Poisson bracket is avoided by the genuine constraints of the
system. This implies that there is no non-commutativity problems
automatically when
one consider the Dirac bracket, in which constraints can be
implemented consistently through variations. Moreover, since the usual
Regge-Teitelboim approach \cite{Regg}, which finds the appropriate
$J[\hat{\xi}]$ such that $\delta_{\eta} J[\hat{\xi}]$ cancels the
boundary part of the variations of the bulk symmetry generator $\delta
H_{\hbox{\scriptsize bulk}} [\hat{\xi}]$, uses the limiting process first and
then the variating,
this approach might not be always true. The only resolution is
that {\it the difference of the two processes is proportional
  to the constraints} as in my case. But since there is no other fundamental
constraints besides the Hamiltonian and momentum constraints in my black
hole system, I am led to a conjecture, {\it for any black hole system}, that\\

{\it Non-commutativity of taking the
limit to go to the horizon and computing variation is proportional to
the Hamiltonian and momentum constraints}.\\

I can not generally prove this
conjecture since it is not clear whether the fall-off conditions that
I have studied in this paper are the most general for the Virasoro
algebra to exist, but this should be the
case for the consistency of functional
differentiation; furthermore, it seems that this conjecture may be
extended to any gauge theories, which have the Gauss' law constraints
as the fundamental constraints, if there are some
appropriate boundaries where the variational principles are well-defined
\cite{Park:99}.

{\it The first term}: The first term in (\ref{delJ}) is the main term. From
(\ref{deln}), one has
\begin{eqnarray}
\delta_{\xi} n^r =-\frac{1}{2 f^2} n^r \delta_{\xi} g_{rr}.
\end{eqnarray}
On the other hand, from (\ref{diff}) one obtains
\begin{eqnarray}
\label{delgrr}
\delta_{\xi} g_{rr} =-2 f^2 (\partial_t
-N^{\hat{\alpha}} \partial_{\hat{\alpha}} )\xi^t +\frac{\beta
f^2}{\pi} (\partial_t -N^{\hat{\alpha}}
\partial_{\hat{\alpha}} )\partial_t \xi^t
+O(N^{-1}),
\end{eqnarray}
where
I have used
$\xi^r=-(N \beta /2 \pi f) (\partial_t -N^{\alpha}
\partial_{\alpha} )\xi^t$  from (\ref{xr(sol)}), $\partial_t
\xi^t=-N^{\hat{\alpha}} \partial_{\hat{\alpha}} \xi^t $ from (\ref{dtxt})
with $N^r=0$, the rotational symmetry
$N^{\alpha} \partial_{\alpha} N^2 =0$, and the condition (\ref{drNbeta}).
Now then, the first term in
(\ref{delJ}) becomes
\begin{eqnarray}
&&-\frac{1}{16 \pi G} \oint_{r_+} d^{n-1} x~ \frac{n^r}{f^2}
\partial_r \hat{\xi_1}^t \delta_2 g_{rr} \sqrt{\sigma}\nonumber \\
&&~~~~=-\frac{1}{16 \pi G} \oint_{r_+} d^{n-1} x ~\sqrt{\sigma}
\xi^t_1 \left(-\frac{2 \beta }{\pi}  N^{\hat{\gamma}}
N^{\hat{\alpha}}
  N^{\hat{\beta}} \partial_{\hat{\gamma}} \partial_{\hat{\alpha}}
  \partial_{\hat{\beta}} \xi^t_2
+ \frac{8 \pi N^{\hat{\alpha}}}{ \beta } \ \partial_{\hat{\alpha}}
\xi^t_2
\right).
\end{eqnarray}
Here, I have taken a safe integration by parts for $\hat{\alpha}$
[$\bar{\alpha}$ part does not contribute] coordinate due to the rotational
symmetry $\partial_{\hat{\alpha}} N=\partial_{\hat{\alpha}}
N^{\hat{\beta}}=0$.

By summarizing the computation,
$\delta_{\xi_2} J[\hat{\xi_1}]$ of (\ref{delJ}) is reduced to
\begin{eqnarray}
\label{delJsum}
\delta_{\xi_2} J[\hat{\xi}_1]
&=&\frac{1}{8 \pi G}  \oint_{r_{+}} d^{n-2} x \left[\delta_{\xi_2} n^r
  \partial_r
\hat{\xi}_{1}^t \sqrt{\sigma}+16 \pi G f^2 \xi^r_1 \delta_{\xi_2} p^{rr} +O(N)
\right] \nonumber \\
&=&\frac{1}{8 \pi G} \oint_{r_+} d^{n-2} x \sqrt{\sigma} \xi^t_1
\left(\frac{\beta }{\pi}  N^{\hat{\gamma}} N^{\hat{\alpha}}
 N^{\hat{\beta}} \partial_{\hat{\gamma}} \partial_{\hat{\alpha}}
  \partial_{\hat{\beta}} \xi^t_2-
\frac{4 \pi N^{\hat{\alpha}}}{ \beta }  \partial_{\hat{\alpha}}
\xi^t_2\right)\nonumber \\
&&~~~~~~+ \oint_{r_+} d^{n-2} x  \hat{\xi_1}^r
\hat{\xi_2}^t {\cal H}^t +O(N).
\end{eqnarray}

On the other hand, notice that
\begin{eqnarray}
J[\{\hat{\xi_1}, \hat{\xi_2}\}_{\hbox{\scriptsize SD}} ]&=&\frac{1}{8 \pi G}  \oint_{r_{+}}
d^{n-2} x~ \left[ n^a D_a
\{\hat{\xi_1}, \hat{\xi_2}\}^t_{\hbox{\scriptsize SD}} \sqrt{\sigma}
+16 \pi G \{\hat{\xi_1},
\hat{\xi_2}\}^a_{\hbox{\scriptsize SD}}p^{r}_a \right] \nonumber \\
&=&\frac{1}{8 \pi G}  \oint_{r_{+}} d^{n-2} x ~\sqrt{\sigma} n^r
\left(\frac{4 \pi f N^{\hat{\alpha}} }{\beta }
  \partial_{\hat{\alpha}}\xi^t_1  \xi^t_2 -(1 \leftrightarrow  2)
\right) +O(N).
\end{eqnarray}
Here I have used the fact that
\begin{eqnarray}
&&\{\hat{\xi_1}, \hat{\xi_2}\}^t_{\hbox{\scriptsize SD}} =2 N
N^{\alpha} \partial_{\alpha} \xi_1^t \xi^t_2 -( 1
\leftrightarrow 2) ,\nonumber \\
&&\{\hat{\xi_1}, \hat{\xi_2}\}^r_{\hbox{\scriptsize SD}} =O(N^2),
~~\{\hat{\xi_1}, \hat{\xi_2}\}^{\alpha}_{\hbox{\scriptsize SD}}
=O(N^2),~~ \{\hat{\xi_1}, \hat{\xi_2}\}^a_{\hbox{\scriptsize SD}}
{p^r}_a =O(N)
\end{eqnarray}
from
(\ref{xr(sol)}), (\ref{drxt(sol)}), (\ref{dtxt}), and $\hat{\xi}^{\alpha}=0$. Then
(\ref{delJsum}) can be written as follows:
\begin{eqnarray}
\delta_{\xi_2} J[\hat[\xi_1]] =J[\{\hat{\xi_1}, \hat{\xi_2}\}_{\hbox{\scriptsize SD}}]
+K[\{\hat{\xi_1}, \hat{\xi_2}\}_{\hbox{\scriptsize SD}}] +\frac{1}{4 \pi}
\oint_{r_+} d^{n-2} x~ \beta N f^{-1}  \{\hat{\xi_1},
\hat{\xi_2}\}^t_{\hbox{\scriptsize SD}} {\cal H}^t +O(N),
\end{eqnarray}
where
\begin{eqnarray}
\label{K(rot)}
K[\{\hat{\xi_1}, \hat{\xi_2}\}_{\hbox{\scriptsize SD}}]&=&\frac{1}{8 \pi G} \oint_{r_+}
d^{n-2} x ~\sqrt{\sigma} \xi^t_1 \left(\frac{\beta }{\pi}
  N^{\hat{\gamma}} N^{\hat{\alpha}} N^{\hat{\beta}}
  \partial_{\hat{\gamma}} \partial_{\hat{\alpha}}
  \partial_{\hat{\beta}} \xi^t_2
+ \frac{4 \pi N^{\hat{\alpha}}}{ \beta }  \partial_{\hat{\alpha}}
\xi^t_2\right)\nonumber \\
&=&\frac{1}{8 \pi G} \oint_{r_+} d^{n-2} x~ \sqrt{\sigma} \xi^t_1
\left(-\frac{\beta }{\pi}  \partial_t \partial_t \partial_t \xi^t_2
- \frac{4 \pi}{ \beta }  \partial_t \xi^t_2\right).
\end{eqnarray}
 From the relation (\ref{RT}), one further has a Virasoro-type algebra
\begin{eqnarray}
\label{sVirasoro}
\{J[\hat{\xi_1}], J[\hat{\xi_2}]\}^* &=&\delta_{\xi_2} J[\hat{\xi_1}]
\nonumber \\
&=&J[\{\hat{\xi_1}, \hat{\xi_2}\}_{\hbox{\scriptsize SD}}]
+K[\{\hat{\xi_1}, \hat{\xi_2}\}_{\hbox{\scriptsize SD}}]
+\frac{1}{4 \pi} \oint_{r_+} d^{n-2} x ~\beta N f^{-1}
\{\hat{\xi_1}, \hat{\xi_2}\}^t_{\hbox{\scriptsize SD}} {\cal H}^t
\end{eqnarray}
with a central term $K[\{\hat{\xi_1}, \hat{\xi_2}\}_{\hbox{\scriptsize SD}}]$. Notice
that, the Virasoro algebra has been generalized to higher dimensions
depending on the number of independent rotations \cite{Myer}: For uncharged black
holes, the number of independent rotations is given by $[(n-1)/2]$,
which is the number of Casimir invariants of the $SO(n-1)$ rotation
group. ($[(n-1)/2]$ denotes the integer part of $(n-1)/2$.) Moreover,
since the relation (\ref{RT}) is valid with the Dirac bracket, where
the constraints are imposed consistently, the last constraint
term in (\ref{sVirasoro}) has no effect on the central term
$K[\{\hat{\xi_1}, \hat{\xi_2}\}^a_{\hbox{\scriptsize SD}}]$.

Now, in order to obtain a more familiar momentum-space Virasoro
algebra, I
adopt the Fourier expansion of (\ref{xtassump}) as usual:
\begin{eqnarray}
\label{Fouriea}
\xi^t=\frac{T}{4 \pi}\sum_{n} \xi_n \mbox{exp} \left\{\frac{2 \pi in}{T}
  \left(t-r_*-N^{\phi}|^{-1}_{r_+} \phi\right)\right\},~
  ~J[\hat{\xi}]=\sum_n \xi_n {\cal J}_n
\end{eqnarray}
with $\xi_n^*=\xi_{-n}$.
The normalization of (\ref{Fouriea}) is chosen such that one can obtain
the standard factor $i (m-n) {\cal J}_{m+n}$ in the Virasoro algebra below.
Contrary to the lower dimensional case ($n\leq 3$), the four
and higher dimensional $\xi_n$ have $x^{\bar{\alpha}}$ dependences in
general, in which the momentum-space representation is different from
the usual one; in this case the central term is expressed as an
integral not just a number. But, for our purpose, I consider
the case where $\xi_n$ has no other coordinate-dependence. Then one
obtains the familiar classical Virasoro algebra in the momentum space
\begin{eqnarray}
\label{mVirasoro}
\{ {\cal J}_m,{\cal J}_n \}^* =-i (m-n) {\cal J}_{m+n} -i K_{m+n},
\end{eqnarray}
where
\begin{eqnarray}
{\cal J}_p=\frac{A_+}{16 \pi G} \frac{T}{\beta} \delta_{p,0},~~
K_{m+n} =\frac{A_+}{16 \pi G} m \left(m^2
-\left(\frac{T}{\beta}\right)^2\right)\frac{ \beta}{T} \delta_{m, -n}
\end{eqnarray}
and $A_+$ is the area of the horizon $r_+$. The standard form of the
central term is also obtained by a constant shift on ${\cal J}_0$
\begin{eqnarray}
\label{L0}
{\cal J}_0 \rightarrow  J_0&=&{\cal J}_0 +\frac{A_+}{32 \pi G} \frac{\beta}{T}
\left(1-\left(\frac{T}{\beta}\right)^2 \right) \nonumber \\
&=&\frac{A_+}{32 \pi G} \frac{\beta}{T}
\left(1+\left(\frac{T}{\beta}\right)^2 \right)
\end{eqnarray}
leading to
\begin{eqnarray}
\label{standard}
\{ J_m,J_n \}^* =-i (m-n) J_{m+n}  -i
\frac{c}{12} m (m^2
-1) \delta_{m, -n}
\end{eqnarray}
with a central charge
\begin{eqnarray}
\label{c}
c=\frac{3A_+}{4 \pi G}\frac{\beta}{T}.
\end{eqnarray} Notice the sign change of the second term in
(\ref{L0}) from the shift.

\begin{section}
{The Virasoro algebra for a non-rotating horizon}
\end{section}

So far, I have considered the Virasoro algebra for the surface
deformation algebra on a black hole horizon when there is at least
one non-zero rotation, i.e., Kerr black hole. But non-rotating
($N^{\alpha}=0$) black hole, which has now a spherical symmetry, can
be analyzed similarly with some modifications in the formulas for the
rotating horizon.

The main differences in the basic relations are as follows
(note $N^r=0$). First, equation (\ref{xr(sol)}) becomes
\begin{eqnarray}
\label{xr(norot)}
\xi^r&=&-\frac{1}{2} \times \frac{N \beta}{\pi f} \partial_t \xi^t
~~(\mbox{non-rotating ~horizon}), \\
\xi^r&=&- \frac{N \beta}{\pi f} \partial_t \xi^t ~~(\mbox{rotating~
  horizon}). \nonumber
\end{eqnarray}

The second difference
is that the condition of fixed angular velocity $\delta_{\xi}
N^{\alpha}|_{r_{+}}=0$ in a grand canonical ensemble produces
\begin{eqnarray}
\xi^{\alpha}& =&\xi^{\alpha} (r, x^{\beta}) ~~(\mbox{non-rotating~horizon}), \\
\xi^{\alpha}& =&-N^{\alpha} \xi^t ~~ (\mbox{rotating~horizon}) \nonumber
\end{eqnarray}
from (\ref{delNalpha}); even without rotation, the angular $Diff$
does not vanish, in contrast to a non-rotating limit of a rotating horizon;
but now there is no connection between $\xi^{\alpha}$ and
$\xi^t$. However, the basic fall-off conditions
(\ref{f=N-1})-(\ref{DN}) and their preserving conditions
(\ref{drsigma})-(\ref{drNbeta}) are the same as for the rotating horizon;
except that some of the
conditions for the non-rotating case are milder than those of the
rotating case, but in this case also, the non-rotating case can be
obtained as a limit of the rotating case;
\begin{eqnarray}
&&\partial_r \xi^{\beta} \leq O(1) ~~(\mbox{non-rotating~horizon}), \\
&&\partial_r \xi^{\beta}+N^{\be} \partial_r \xi^t \leq O(1)
~~(\mbox{rotating~horizon}) \nonumber
\end{eqnarray}
is one example.

The remaining analysis on the Virasoro algebra is
straightforward and it is easily found that the coordinate-space
Virasoro algebra has the form of (\ref{sVirasoro}) with
\begin{eqnarray}
J[\hat{\xi} ]&=&\frac{1}{2}\times \frac{1}{8 \pi G}  \oint_{r_{+}}
d^{n-2} x~\sqrt{\sigma}  n^a D_a
\hat{\xi}^t, \nonumber \\
\label{K(norot)}
K[\{\hat{\xi}_1, \hat{\xi}_2 \}_{\hbox{\scriptsize SD}}]&=&\frac{1}{2} \times \frac{1}{8
    \pi G} \oint_{r_+} d^{n-2} x \sqrt{\sigma} \left( \frac{4
    \pi}{\beta}  \partial_t \xi^t_{1
    } \xi^t_{2} +\frac{\beta}{\pi}\partial_t \xi^t_{1 }
    \partial_t \partial_t \xi^t_{2 }
-2 \partial_t \xi^t_{1} \partial_t \xi^t_{2}
  -2\partial_t \partial_t \xi^t_{1} \xi^t_{2} \right), \\
\{\hat{\xi}_1,\hat{\xi}_2 \}^t_{\hbox{\scriptsize SD}}&=&\frac{1}{2}\times [-2 N \partial_t
    \xi^t_1 \xi^t_2 -(1 \leftrightarrow 2) ]. \nonumber
\end{eqnarray}
The overall factor $1/2$ is the result of the same factor in
(\ref{xr(norot)}). But, due to the absence of the relation
(\ref{dtxt}), this is not truly a Virasoro algebra because
the derivative $\partial_t$ cannot be integrated by parts and so the
manifest $1\leftrightarrow 2$ antisymmetry is not attained in contrast to
the generic definition (\ref{RT}). Hence, in order that $J[\hat{\xi}]$
be a symmetry generator as (\ref{RT}), I am required to connect
$\partial_t \xi^t$ with $\partial_{\hat{\alpha}} \xi^t$ for some
angular coordinates $x^{\hat{\alpha}}$ as follows\footnote{On the other hand,
  with no angular dependence, the orthogonality
disappear, and the whole
algebra contains the time-dependent factor $\mbox{exp} \{2 \pi i
  (m+n)(t-r_*)/T\}$ as well as other unwanted terms which are proportional to
 $(m^2-n^2)$; This is the same situation as in two-dimensional gravity
 \cite{Park:00}.}
:
\begin{eqnarray}
\partial_t \xi^t=v^{\hat{\alpha}} \partial_{\hat{\alpha}} \xi^t.
\end{eqnarray}
But now, the speed $v^{\hat{\alpha}}$, which is arbitrary, has no
connection to the horizon's rotation. The last two terms
cancel from the integration by parts and the central term
(\ref{K(norot)}) becomes the same as (\ref{K(rot)}) with the differences in
the over-all factor $1/2$ and with $v^{\alpha}$ instead of
$N^{\alpha}$:
\begin{eqnarray}
K[\{\hat{\xi_1}, \hat{\xi_2}\}^a_{\hbox{\scriptsize SD}}]&=&\frac{1}{2} \times \frac{1}{8
  \pi G} \oint_{r_+} d^{n-2} x ~\sqrt{\sigma} \xi^t_1 \left(\frac{\beta
    }{\pi}  v^{\hat{\gamma}} v^{\hat{\alpha}} v^{\hat{\beta}}
  \partial_{\hat{\gamma}} \partial_{\hat{\alpha}}
  \partial_{\hat{\beta}} \xi^t_2
+ \frac{4 \pi v^{\hat{\alpha}}}{ \beta }  \partial_{\hat{\alpha}}
  \xi^t_2\right)\nonumber \\
&=&\frac{1}{2}\times \frac{1}{8 \pi G} \oint_{r_+} d^{n-2} x~
  \sqrt{\sigma} \xi^t_1 \left(-\frac{\beta }{\pi}  \partial_t
  \partial_t \partial_t \xi^t_2
- \frac{4 \pi}{ \beta }  \partial_t \xi^t_2\right).
\end{eqnarray}
Then, for the $Diff$ $\xi^t$ which ``lives'' at the horizon
\begin{eqnarray}
\xi^t =\xi^t \left[ t-r_* +\frac{1}{v^{\hat{\alpha}}} x^{\hat{\alpha}} \right]
\end{eqnarray}
the momentum-space Virasoro algebra has the standard form (\ref{standard}) with
\begin{eqnarray}
\label{L0(norot)}
J_0 =2 \times \frac{A_+}{32 \pi G} \frac{\beta}{T}
\left(1+\left(\frac{T}{\beta}\right)^2 \right), ~~
c=2 \times \frac{3A_+}{4 \pi G}\frac{\beta}{T}.
\end{eqnarray}
The change of factor $1/2$ in (\ref{K(norot)}) to $2$ in
(\ref{L0(norot)}) comes
from the normalization of the Fourier expansion of $\xi^t$
\begin{eqnarray}
\xi^t=2 \times \frac{T}{4 \pi}\sum_n \xi_n \mbox{exp}\left\{\frac{2 \pi in}{T}
  \left(t-r_*-{v^{\phi}} ^{-1}\phi \right)\right\}
\end{eqnarray}
in order to obtain the correct standard form factor $i(m-n) J_{m+n}$
in (\ref{mVirasoro}). Notice that furthermore, since $\phi$ should be
periodic for $2 \pi$ rotation, $v^{\phi}$ behaves as $2 \pi/T$. \\

\begin{section}
{Canonical quantization and black hole entropy}
\end{section}

The computation of black hole entropy from the canonical
quantization of the classical Virasoro algebra is rather
straightforward at this stage.

When one uses the standard canonical quantization rule
\begin{eqnarray}
[L_m, L_n ]=i \hbar \{J_m, J_n \}, ~~J_{m+n} \rightarrow L_{m+n},
\end{eqnarray}
where $L_m$ is a quantum operator, the classical Dirac bracket algebra
(\ref{standard}) becomes an operator algebra
\begin{eqnarray}
\label{{L,L}}
[L_m,L_n]=\hbar (m-n) L_{m+n}+ \frac{\hbar c}{12} m(m^2-1) \delta_{m,-n}.
\end{eqnarray}
By considering the transformation
\begin{eqnarray}
\label{Lhat}
L_m \rightarrow \hbar (: \hat{L}_m: +\hbar a \delta_{m,0} ),
\end{eqnarray}
(\ref{{L,L}}) becomes the standard operator Virasoro algebra for the
normal ordered operator $:\hat{L}_m:$ ($a$ is some number)
\begin{eqnarray}
[:\hat{L}_m:,:\hat{L}_n:]=(m-n) :\hat{L}_{m+n}: +\frac{c_{\hbox{\scriptsize tot}}}{12}
m(m^2-1) \delta_{m,-n}
\end{eqnarray}
with
\begin{eqnarray}
\label{ctot}
c_{\hbox{\scriptsize tot}}=\frac{c}{\hbar} +c_{\hbox{\scriptsize quant}}.
\end{eqnarray}
(See also Ref. \cite{Carl:00} for some related discussions.)
Here, $c_{\hbox{\scriptsize quant}}$, which is $O(1)$, is the quantum effect due to operator
reordering.

With the Virasoro algebra of $:\hat{L}_m:$ in the standard form,
which is defined on the {\it plane}, one can use Cardy's formula for the
asymptotic states \cite{Card,Carl:95,Carl:98,Carl:99''}
\begin{eqnarray}
\mbox{log} \rho (\hat{\Delta}) \sim 2 \pi \sqrt{ \f{1}{6} \left(c_{\hbox{\scriptsize tot}}-24
  \hat{\Delta}_{\hbox{\scriptsize
  min}}\right)\left(\hat{\Delta}-\frac{c_{\hbox{\scriptsize tot}}}{24}\right) },
\end{eqnarray}
where $\hat{\Delta}$ is the eigenvalue, called conformal weight, of
$:\hat{L}_0:$ for a black
hole quantum state $|$black hole$\rangle$ and
$\hat{\Delta}_{\hbox{\scriptsize min}}$ is its minimum value. When this is expressed in
terms of the classical eigenvalue $\D \equiv J_0$ and the central charge $c$
through
\begin{eqnarray}
\hat{\Delta}=\f{\D}{\hbar}- \hbar a
\end{eqnarray}
from (\ref{Lhat}) and (\ref{ctot}), one obtains
\begin{eqnarray}
\label{Cardy}
\mbox{log} \rho (\D) &\sim &\frac{2 \pi}{\hbar} \sqrt{
 \f{1}{6} \left(c-24
 \Delta_{\hbox{\scriptsize min}}+\hbar c_{\hbox{\scriptsize quant}}+24 \hbar^2
 a\right)\left(\Delta-\frac{c}{24}-\hbar^2 a -\frac{\hbar
 c_{\hbox{\scriptsize quant}}}{24}\right) } \nonumber \\
&=&\frac{2 \pi}{\hbar} \sqrt{ c_{\hbox{\scriptsize eff}}
 \Delta_{\hbox{\scriptsize eff}}/6 +O(\hbar) }
\end{eqnarray}
with
\begin{eqnarray}
c_{\hbox{\scriptsize eff}}=c-24\Delta_{\hbox{\scriptsize min}},~~
\Delta_{\hbox{\scriptsize eff}}=\Delta -\frac{c}{24}.
\end{eqnarray}
This result shows explicitly how the classical Virasoro generator and
central charge can give the correct order of the semiclassical BH entropy
$(c=k=1)$
\begin{eqnarray}
\label{BH}
S_{\hbox{\scriptsize BH}}\sim\frac{A_+}{4 \hbar G}
\end{eqnarray}
since $c\sim A_+/G$ and $\Delta\sim A_+/G$; details on the
numerical factor $1/4$ of the BH entropy depends on
$\Delta_{\hbox{\scriptsize min}}$, which has to be put in $by~ hand$, and the
quantum correction due to reordering gives negligible $O(1/\sqrt{\hbar})$ effect to
the entropy when one considers the macroscopic ensemble of $c_{\hbox{\scriptsize eff}}
\Delta_{\hbox{\scriptsize eff}} \gg 1$.

Now, let me compute the black hole entropy explicitly. For
rotating horizons,
(\ref{Cardy}) produces the entropy
\begin{eqnarray}
S=\mbox{log} \rho( \Delta ) \sim \frac{2 \pi}{\hbar}
\sqrt{\left(\frac{A_+}{16 \pi G}  \right)^2-\frac{A_+}{8 \pi G}
  \frac{T}{\beta} \D_{\hbox{\scriptsize min}} }
\end{eqnarray}
with
\begin{eqnarray}
\Delta_{ \hbox{\scriptsize eff}}= \frac{A_+}{32 \pi G} \frac{T}{\beta}.
\end{eqnarray}
This gives the BH entropy (\ref{BH}) if one takes
\begin{eqnarray}
\label{L0min}
\Delta_{\hbox{\scriptsize min}}=-\frac{3 A_+}{32 \pi G} \frac{\beta}{T}
\end{eqnarray}
such as
\begin{eqnarray}
c_{\hbox{\scriptsize eff}} = \frac{3 A_+}{ \pi G} \frac{ \beta}{T}.
\end{eqnarray}
Note that $\D_{\hbox{\scriptsize min}}$ has an explicit $\beta/T$
dependence.
 But from the
fact that this value is outside of the classical spectrum of
$\Delta \geq (16 \pi G)^{-1} A_+ $, one can expect that the ground
state of this black hole may be described by another class
of black holes; or this might be relevant to the non-commutative
spacetime near the ground state black hole which, probably, is very light
\cite{Park:01}.

On the other hand, for non-rotating horizons, there is an additional
factor ``$2$'' in (\ref{L0(norot)}) and this has a
remarkable consequence for the entropy computation: The entropy from
the Cardy formula gives
\begin{eqnarray}
S\sim \frac {2 \pi}{\hbar} \sqrt{ \left( \frac{A_+}{ 8 \pi G
    }\right)^2-\frac{A_+}{4 \pi G}  \frac{T}{\beta} \Delta_{\hbox{\scriptsize min}} }
\end{eqnarray}
with
\begin{eqnarray}
\Delta_{\hbox{\scriptsize eff}}= \frac{A_+}{16 \pi G} \frac{T}{\beta}.
\end{eqnarray}
 This gives the BH entropy (\ref{BH}) with a $T$-independent ground state
\begin{eqnarray}
\Delta_{\hbox{\scriptsize min}} =0
\end{eqnarray}
such as
\begin{eqnarray}
c_{\hbox{\scriptsize eff}} = \frac{3 A_+}{2 \pi G} \frac{\beta}{T}.
\end{eqnarray}
This means that for an arbitrary choice of $T$ and hence
$v^{\hat{\alpha}}$ the entropy is uniquely defined. Moreover, the
arbitrary $T$ dependences in $c$ and $\Delta$  exactly cancel
each other and one obtains
the $T$-independent correct entropy\footnote{This was
  claimed even in the rotating case by Carlip \cite{Carl:99} due to
  the additionally introduced factor $2$, which is unclear in the
  present context. But his claim is exactly
  realized in the case of non-rotating case or in the
  $c_{\hbox{\scriptsize eff}}$ and $\Delta_{\hbox{\scriptsize eff}}$
  even when there is rotation. The
  connection to the usual $T=\beta$ relation in the path integral
  formulation \cite{Gibb',Brow:93} needs further studies.}.

The appropriate ground state is different for the rotating and the
non-rotating black holes, but otherwise it has a $universality$
for a wide variety of other black holes: The fall-off conditions
described in section 3 involve only weak assumptions about the
black hole, and once $\Delta_{min}$ is fixed, say for an isolated,
uncharged Kerr black hole, its value is determined also for a
large number of other black holes carrying electric or magnetic
charge (details will appear elsewhere
\cite{Park:02b}).\\

 \begin{section}
{Applications}
\end{section}

I have shown that the statistical entropies of rotating and
non-rotating stationary $(N^r=0)$ black holes through the Cardy formula
have a universal form if the fall-off conditions and several
fall-off-preserving
conditions are satisfied. So, the problem of the entropy
computation is reduced to verification of the
conditions. My considerations are general enough that
almost all the known solutions satisfy the required conditions. In this section I
briefly discuss some of these. I shall adopt
different units of $G$ depending on the usual conventions for the
solutions in the literatures.\\

\begin{subsection}
{The Kerr and Schwarzschild black holes}
\end{subsection}
The Kerr black hole solution, with one-rotation, in the 4 and higher
dimensions is given, in the standard form (\ref{ds2})
\cite{Myer,Chan} as
\begin{eqnarray}
\label{Kerr}
d s^2 =-\frac{\rho^2 \delta}{\Sigma^2} dt^2
+\frac{\Sigma^2}{\rho^2} \mbox{sin}^2 \theta \left(d \phi-\frac{\mu
    a}{r^{n-5} \Sigma^2} dt \right)^2
+\frac{\rho ^2}{\delta} d r^2 + \rho^2 d \theta^2 +r^2 \mbox{cos}^2 \theta
d \Omega^{n-4},
\end{eqnarray}
where
\begin{eqnarray}
\rho^2 =r^2 +a^2 \mbox{cos}^2 \theta, ~~\delta=r^2 + a^2 -\frac{\mu} {r^{n-5}},
 ~~\Sigma^2=\rho^2 (r^2 +a^2) +\frac{\mu}{r^{n-5}} a^2 \mbox{sin}^2 \theta
\end{eqnarray}
and $\Omega^{n-4}$ is the line element on a unit $n-4$ sphere;
$\mu$ and $a$ are the mass and the angular momentum parameters
respectively. One can easily check that all the fall-off,
fall-off-preserving, and differentiability conditions are
satisfied in the $n=4$ non-extremal Kerr-solution such that it has
the universal statistical entropy (\ref{Cardy}) and gives the BH
entropy (\ref{BH}) with $A_+=4 \pi (r_+^2 +a^2 )=8 \pi M r_{+}$,
and
\begin{eqnarray}
\Delta_{\hbox{\scriptsize min}} =-\frac{3 \pi}{T}
\frac{M^2(M+\sqrt{M^2-a^2})^2}{\sqrt{M^2-a^2}}.
\end{eqnarray}
Here, $M$ and $a$ are the ADM mass and the angular momentum per unit mass
of the black hole, respectively and $r_{\pm}=MG \pm \sqrt{M^2 -a^2}~(G=1)$.
The higher dimensional solutions have exactly the same results if
only the one-rotation solution (\ref{Kerr}) is concerned.

Moreover, since the Schwarzschild solution can be obtained as
a non-rotating limit of the Kerr solution, it is a trivial matter to
check that the
Schwarzschild solution also satisfies all the fall-off and other related
conditions.
So, in this case also the universal entropy form can be applied.
Note that, the ground state of the non-rotating black hole is
contained in the full spectrum of $\D^{a=0}$ and obtained as
$M\rightarrow 0$ limit, which is nothing but the flat spacetime:
$\Delta^{a=0}=((8 \pi M^3/T) +(MT/8 \pi))\rightarrow 0$.

The lower ground state for the
rotating black hole compared to the non-rotating one may be
understood qualitatively as follows:
Let me consider a black hole with a tiny angular velocity, which is almost
static. Then, increase the angular speed adiabatically slowly such as
no other physical properties of the black hole are changed. But in
order that the horizon is not naked by this adiabatic process, the
mass of the black hole should also be simultaneously increased by
$\delta M^2 \geq \delta a^2$. In order to accommodate this increase
in mass, the vacuum becomes lower without changing the identity of
the black hole. But a complete understanding still needs to be
discovered.\\

\begin{subsection}
{With a cosmological constant}
\end{subsection}

My analysis can be also generalized to include the cosmological
constant(CC) term with only a small modification in the formulas of the
preceding sections.

The CC term
\begin{eqnarray}
S_{CC}=-\frac{1}{16 \pi G} \int d^{n} x N \sqrt{h} (2 \Lambda )
\end{eqnarray}
added to the action (\ref{SEH}) changes the Hamiltonian constraints as
\begin{eqnarray}
{\cal H}^t =-\frac{\sqrt{h}}{16 \pi G}( R - 2 \Lambda)
+\frac{16 \pi G}{\sqrt{h}}
\left(p_{ab} p^{ab} -\frac{1}{n-2} p^2 \right).
\end{eqnarray}
But since this additional term does not generate any surface term in
the variations, almost all the results of the boundary conditions and the
surface deformations in the preceding sections are unchanged. The only
exception is the
computation of (\ref{fifth}), which produces the constraints term in
the Virasoro algebra (\ref{sVirasoro}). But in this case again, there
is no effect of $\Lambda$ at the horizon: From
\begin{eqnarray}
{\cal H}^t_{\Lambda}={\cal H}^t_{\Lambda=0} +\frac{\sqrt{h}}{8 \pi G} \Lambda
                    \cong {\cal H}_{\Lambda=0} ^t, ~~
\delta_{\xi} p^{rr}_{\Lambda} =\delta_{\xi}
p^{rr}_{\Lambda=0}-\hat{\xi}^t \frac{\sqrt{h}}{16 \pi G} h^{rr} \Lambda
\cong \delta_{\xi} p^{rr}_{\Lambda=0},
\end{eqnarray}
one finds that  the variations of $\delta_{\xi_2}
J_{\Lambda}[\hat{\xi_1}]$ involving $\delta_{\xi} p^{rr}_{\Lambda}$
is again the constraint term, which is the same as (\ref{fifth}).

However, if
the CC generates its own horizon which  is not due to the
black hole, it can be treated as another independent application of
my original method. The interesting examples are the BTZ solution in
$n=3~(\Lambda <0)$ and rotating de-Sitter space solutions $(\Lambda
>0)$. \\

\begin{subsubsection}
{The BTZ black hole}
\end{subsubsection}

The BTZ black hole solution \cite{BTZ} in $n=3~(\Lambda <0)$ is
similar to the Kerr solution. But I consider this example since there
are some points which are worthy of studying  in comparison with the
other entropy
computations in different contexts.

The BTZ solution is given by the
standard form (\ref{ds2}) with $(G=1/8)$
\begin{eqnarray}
f^{-2}=N^2=\frac{(r^2-r^2_{+} ) (r^2-r^2_{-})}{r^2}, ~~N^r=0,
~~N^{\phi}=\frac{r_{+} r_{-}}{r^2},~~N_{\phi}=r_{+} r_{-}, ~~\sigma_{\phi \phi}=r^2.
\end{eqnarray}
All the subsidiary conditions are satisfied the same way as in the
Kerr-solution and so, the universal entropy formula (\ref{Cardy}) is also
applicable here with
$A_+=2 \pi r_{+}$ and
\begin{eqnarray}
\Delta_{\hbox{\scriptsize min}} =-\frac{3 \pi l^2}{T} \frac{r_{+}^2}{r_{+}^2
  -r_{-}^2} ~~(J \neq 0), ~~\Delta_{\hbox{\scriptsize min}} =0~~(J=0).
\end{eqnarray}
Here, $r_{\pm}=(l/\sqrt{2}) \sqrt{M \pm \sqrt{M^2-(J/l)^2}}$. The
ground state of non-rotating solution is obtained as $M\rightarrow
0_{-}$: $\lim_{M\rightarrow 0_{-}}\Delta^{J=0}=-(2 \pi)^{-1}|M|T =
0$. This is contrary to Strominger's entropy computation, in which
the Virasoro algebra is an algebra for an asymptotic infinity, not
for the horizon and the ground state gives $M=-1$.  On the other
hand, the factor ``$2$'' in the non-rotating case
(\ref{L0(norot)}) corresponds to the ``two'' copies of the
Virasoro algebra for the isometry group at spatial
infinity \cite{Brow:86,Stro:98,Bala:99}.\\

\begin{subsubsection}
{The rotating de-Sitter space:~$n=3$ (Kerr-dS$_3$)}
\end{subsubsection}

The de-Sitter space of $\Lambda >0$ is peculiar in that it has its own
horizon, called cosmological horizon, without black holes
\cite{Gibb,Dese,Mald,Park:98,Stro:01}. Moreover, remarkably this space
can have a $rotating$ horizon with no lower bound of the mass, which
is contrary to the Kerr and the BTZ
solutions. The $n=3$ solution was studied first in Ref. \cite{Park:98} and
more recently studied in other contexts in Refs. \cite{Bala,Cunh}. The
rotating de-Sitter solution without black holes exists also for $n\geq
4$ and can be simply obtained from a zero-mass-black-hole limit in the
usual Kerr-de Sitter black hole solution. Moreover, the computation in
my method is interesting because the previous computations in the
$n=3$ case
which have used the isometry at spatial infinity, which is hidden inside
the horizon, produce  complex $\Delta$ (and imaginary $c$ for the
Chern-Simons formulation) when rotations are
involved \cite{Park:98,Bala,Cunh} always. But strangely enough, the final
result of the entropy is real valued and identical to the usual BH
entropy \cite{Gibb}. So, it is important
to check whether the complex number disappears in all the
intermediate steps or not when I treat the correct boundary $r_+$ not the
suspicious boundary at infinity. In this subsection, I first consider
$n=3$ solution and then consider $n\geq 4$ in the next subsection.

The rotating de-Sitter in $n=3, \Lambda=l^{-2}>0$ is given by
the standard metric (\ref{ds2}) with $(G=1/8)$
\begin{eqnarray}
f^{-2}=N^2=M-\left(\frac{r}{l}\right)^2+\frac{J^2}{4 r^2},~~N^r=0,
~~N^{\phi}=-\frac{J}{2 r^2}, ~~N_{\phi}=-\frac{J}{2}, ~~\sigma_{\phi \phi}=r^2,
\end{eqnarray}
where $M$ and $J$ are the mass and angular momentum parameters of the
solution, respectively. The solution has only one horizon at
$
r_{+}=l/\sqrt{2} \sqrt{M +\sqrt{M^2 +(J/l)^2}}.
$
Notice that there is no constraint on $M^2$ bounded by $J^2$ for a
horizon to exit: Even a negative value of $M$ is allowed when $J\neq
0$ such as the mass spectrum (ranging from
$-\infty$ to $\infty$) is continuous and so there is no mass gap in contrast
to the BTZ solution; for $J=0$  case, there is no horizon for $M<0$.

The fall-off forms of the metric are the same as the BTZ solution
and so the entropy also has the universal from as (\ref{Cardy})
with $A_+=2 \pi r_+$ and
\begin{eqnarray}
\Delta_{\hbox{\scriptsize min}}=-\frac{3 \pi}{T \Lambda}~~(J\neq0),
~~\Delta_{\hbox{\scriptsize min}}=0~~(J=0).
\end{eqnarray}

The static de-Sitter ground state solution is obtained as $M\rightarrow 0_{-}$:
$\lim_{M\rightarrow 0_{-}}\Delta^{J=0}=(2 \pi)^{-1} T l^2  |M|
=0$. Note that there is no complex number at any step of computation.
\\

\begin{subsubsection}
{The rotating de-Sitter solution:~$n\geq 4$ (massless Kerr-dS$_n$)}
\end{subsubsection}

The rotating de-Sitter solution for $n\geq 4$ is obtained from the
$M=0$ reduction of the Kerr-de Sitter solution
\cite{Gibb,Hawk:98}. In the standard form (\ref{ds2}) with ($n=4$
for simplicity; but it is a trivial matter to generalize to higher
dimensions), the solution is given by $(G=1)$
\begin{eqnarray}
&&N^2=\frac{\Delta_r \Delta_{\theta}}{(r^2 +a^2)
  \Xi^3},~~f^2=\frac{\rho^2}{\Delta_r},
  ~~N^r=0,~~N^{\theta}=0,~~N^{\phi}=-\frac{1}{3} \frac{\Lambda
  a}{\Xi},\nonumber \\
&&N_{\phi}=-\frac{1}{3} \frac{\Lambda a (r^2 +a^2)}{\Xi^2} \mbox{sin}^2 \theta ,
~~\sigma_{\phi \phi} =\frac{(r^2 +a^2)}{\Xi} \mbox{sin}^2 \theta,
~~\sigma_{\theta \theta}=\frac{\rho^2}{\Delta_{\theta}}
\end{eqnarray}
with
\begin{eqnarray}
\rho^2 =r^2 +a^2 \mbox{cos}^2 \theta, ~~\Xi=1+\frac{\Lambda a^2}{3}
,~~\Delta_r=(r^2 + a^2)(1- \frac{\Lambda r^2} {3}),
~~\Delta_{\theta}=1+\frac{\Lambda a^2}{3} \mbox{cos}^2 \theta,
\end{eqnarray}
which has the cosmological horizon at $r_+=\sqrt{3/\Lambda}$.

This is very similar to the Kerr-solution and so one can easily check
that all the conditions of the Kerr-solution are still satisfied. So the
universal entropy (\ref{Cardy}) can be applied in this
case, with $A_+=4 \pi r_{+}^2$ and
\begin{eqnarray}
\Delta_{\hbox{\scriptsize min}}=-\frac{3 \pi r_{+}^3}{4 T} ~~(a \neq 0),
~~\Delta_{\hbox{\scriptsize min}}=0 ~~(a=0).
\end{eqnarray}
The ground sate of non-rotating de-Sitter space is obtained
$\Delta^{a=0}=r_{+} ((\pi r_{+}^2/2 T) +(T/8 \pi))
\rightarrow 0$ as $r_+=\sqrt{3/\Lambda} \rightarrow 0$, which
corresponds to giving an infinite moment of inertia to the space. \\

\begin{section}
{Concluding remarks }
\end{section}

I have shown that almost all the known solutions with the horizon have the
universal statistical entropy.  This is identical to the
Bekenstein-Hawking entropy when the appropriate ground states are
chosen and the
higher order quantum corrections of operator orderings are neglected. Here the
existence of the classical Virasoro algebra at the horizon was
crucial to the results. The remaining questions are as follows. \\

1. How can we understand extremal black holes in my method ? Can this method
   explain the discrepancy between the gravity side
   \cite{Hawk:94}\footnote{There is a debate on the gravity side
   also. See Ref. \cite{Zasl}.} and the string theory side
   \cite{Stro:96}, which claim different entropies for the extremal
   black hole ? \\

2. Can my method be generalized to non-stationary metrics such that
    expanding or collapsing horizons can be treated \cite{Brun} ? \\

3. Can gauge and matter fields be introduced without
   perturbing the universal statistical entropy formula ? Can this
   study give another proof of the no-hair conjecture ?

Some of the questions are being studied and will appear elsewhere \cite{Park:02b}.\\

Finally, computing the symplectic structure on the constraint surface
${\cal H}_{\mu} \approx 0$ with a horizon boundary through the Dirac bracket
method or the symplectic reduction will be an outstanding challenge. The higher
order quantum corrections of black hole entropy can be computed by
quantizing the classical symplectic structure.\\
\begin{center}
 {\bf Acknowledgment}
\end{center}

I would like to thank Dimitra Karabali, Parameswaran Nair, Peter
Orland, Alexios
Polychronakos, Bunji Sakita and Stuart Samuel for warm
hospitality while I have been staying at City College. I would like to thank
Steve Carlip and Gungwon Kang for the useful discussions in the very beginning
of this work and Yun Soo Myung for inviting me to Inje University,
where some of the
early ideas were initiated. I thank to Roman Jackiw for introducing
the  interesting world of classical Virasoro algebra while I stayed at
MIT and continuous hospitality after then. This work was supported in
part by postdoctoral fellowships program from
Korea Science and Engineering Foundation (KOSEF) and in part by a CUNY
Collaborative Incentive Research Grant.

\appendix

\section{Analysis on the boundary ${\cal C}$}

The variation in $H'[N, N^a]$ due to arbitrary variations in $h_{ab},n^a,N,N^a$is
%
\begin{eqnarray}
\label{delH'}
\delta H' [N, N^a]
&=& \mbox{bulk terms} +\frac{1}{8 \pi G} \int _{\cal C} d^{n-2} x
\left[ \frac{16 \pi G}{2-n} \sqrt{\frac{\sigma}{h}} n^r N_r \delta p
\right. \no\\
&&\left. +
\delta n^r \left(\partial_r N \sqrt{\sigma}
+\frac{16 \pi G}{2-n} \sqrt{\frac{\sigma}{h}} N_r p \right)
-\frac{16 \pi G}{2(2-n)}\sqrt{\frac{\sigma}{h}} n^a n^b n^r N_r p \delta
h_{ab} \right. \no\\
&&\left. + \delta N^a \left(16 \pi G {p_a}^r
+ \frac{16 \pi G}{2-n} \sqrt{\frac{\sigma}{h}} n_a p \right)
+\frac{16 \pi G}{2-n} \sqrt{\frac{\sigma}{h}} \delta h_{ra} n^r N^a p
 \right]
\eeq
Here, I have also assumed the boundary conditions
(\ref{Nbound}) and (\ref{sbound}) in order to remove the
problematic terms which persist also on ${\cal C}$ as in
(\ref{delH}).

The boundary contributions to the bulk equation of motion are
\begin{eqnarray}
\label{{hab,H'}}
&&\{ h_{ab} (x), H'[N, N^a] \} |_{\hbox{\scriptsize boundary}} =\delta
(r-r_{\cal C}) \left(\frac{2}{2-n }\sqrt{\frac{\sigma}{h}} n^r N_r
  h_{ab}\right), \\
\label{{pi,H'}} &&\{p^{ab} (x), H'[N, N^a]\} |_{\hbox{\scriptsize
boundary}} =-
 \delta (r-r_{\cal C}) \left[\sqrt{\frac{\sigma}{h}} \frac{2}{2-n}  n^r
 N_r p^{ab} \right. \\
&&~~~~\left. +\frac{2}{2-n} \sqrt{\frac{\sigma}{h}} \left(
 {\delta^{\left(a \right.}}_r N^{\left. b \right)} n^r
-\frac{1}{2} n^a n^b n^r N_r-\frac{1}{2} n^{\left( a \right.}
 h^{\left. b \right) r} N_r \right) p
 -\f{\sqrt{\sigma}}{16 \pi G} n^{\left(a \right.} h^{\left. b \right) r} \partial_r N
  \right].\nonumber
\end{eqnarray}
The  boundary contribution to the momentum constraints is
\begin{eqnarray}
\label{H'abound}
{\cal H'}_a |_{\hbox{\scriptsize boundary}} = \delta (r-r_{\cal C})
\left(\frac{2}{2-n}\sqrt{\frac{\sigma}{h}} n^r h_{ra} p + 2{p^r}_a \right)
\end{eqnarray}
In order to achieve the
extremality under the variations with the boundary contributions, I
need the boundary conditions on ${\cal C}$ such that all the terms of
$\delta h_{ab}, \delta p^{ab}, \delta N^a$ vanish which would be
impossible in general. This can be easily checked within our black hole setup,
which shows non-vanishing boundary contributions,
$
\{ p^{r \alpha} (x), H'[N, N^a]\} |_{\hbox{ \scriptsize boundary} }  \leq O (1),~
\{ p^{\alpha \beta} (x), H'[N, N^a] \} |_{\hbox{ \scriptsize boundary}}  \leq O (1),~
\{ h_{rr} (x), H'[N, N^a]\} |_{\hbox{ \scriptsize boundary} }  \leq O(1),~
{\cal H'}_r |_{ \hbox{\scriptsize boundary} }  \leq O (N^{-2})$.

Finally, I note that there is a special choice of
the slicing, called {\it maximal slicing} \cite{Dira,ADM,Regg}, $p=0$, i.e.,
  $p^{\al \be}=0$ (but ${p^r}_{\alpha}=O(1)$ needs
  not be zero), in which the Hamiltonian $H'[N, N^a]$ is well defined,
  i.e., there is no boundary contribution,
when the coordinate system $N_r=0$
    is used. But, even in this case, the $Diff$ generator
    $H'[\hx^t, \hx^a]$ is still ill-defined: $\{h_{rr} (x), H'[\hx^t,
    \hx^a]\} |_{\hbox{\scriptsize boundary}}
= \delta ( r-r_+) ((2-n) \sqrt{h})^{-1} 2 \sqrt{\sigma} n_r \xi^r~ h_{rr}
\leq O(1) $ \cite{Park:99'}.

\section{Computing $2 \oint _{r_+} d^{n-2} \hat{\xi_1}^r \delta_2
{p^r}_r$}

In this Appendix, I compute $2 \oint _{r_+} d^{n-2}
\hat{\xi_1}^r \delta_2 {p^r}_r $ of (\ref{fifth}). From
(\ref{delhp}), one has
\beq
\label{delpi(App)}
\de_\xi {p}^{ab} &=&\{p^{ab}, H[\hx]\}\nonumber \\
 &=&-\hx^t \f{\sqrt{h}}{16 \pi G} \left(R^{ab}-\frac{1}{2} h^{ab} R\right) +
 \f{\sqrt{h}}{16 \pi G} (D^a D^b \hx^t-
h^{ab} D_c D^c \hx^t) \nonumber \\
&&+\frac{8 \pi G \hx^t}{ \sqrt{h}} h^{ab} \left(p_{cd} p^{cd} -\frac{1}{n-2} p^2
 \right) -\frac{32 \pi G \hx^t}{\sqrt h} \left({p^b}_c p^{ac}
   -\frac{1}{n-2} p^{ab} p
\right) \nonumber \\
&&+D_c (\hx^c p^{ab}) -D_c (\hx^a p^{bc}) -D_c (\hx^b p^{ac} ).
\eeq
In order to determine what is involved in the computation of
 (\ref{fifth}), let me expand the integrand of (\ref{fifth}) as
\beq
\label{delpirr(App)}
\hx^r_1 \de_2 {p_r}^r =\hx^r_1 f^2 \de_2 p^{rr} +\hx^r_1 \de_2
 h_{\al r} p^{\al r} +\hx^r_1 \de_2 h_{rr} p^{rr}
\eeq
by writing ${p_r}^r=h_{br} p^{br}$. [Here and after, I do not
 restrict to the case of $N^r=0$ for generality, though I consider
 $\hx^\al=0$ to
 avoid unnecessary complication.] The second and the third terms are
 definitely $O(N^2)$ and $O(N)$,
respectively. $Naively$, the first term is $O(N)$ if
 the fall-off condition (\ref{pirr}) is preserved by $Diff$. But this
 is not a trivial matter. Let me compute this first term in detail.

With the help of (\ref{delpi(App)}), the first term of (\ref{delpirr(App)})
becomes
\beq
\label{delpiall(App)}
\hx^r_1 f^2 \de_2 p^{rr}&=&A+B+C+D,\no\\
A&=&-f^2 \hx^r_1\hx^t_2 \f{\sqrt{h}}{16 \pi G} \left(R^{rr}-\frac{1}{2} h^{rr}
  R\right), \no\\
B&=&f^2 \hx^r_1 \f{ \sqrt{h}}{16 \pi G} (D^r D^r \hx^t_2-
 h^{rr} D_c D^c \hx^t_2), \nonumber \\
C&=&\frac{8 \pi G f^2 \hx^r_1 \hx^t_2}{ \sqrt{h}} h^{rr} \left(p_{cd} p^{cd}
-\frac{1}{n-2} p^2
 \right) -\frac{32 \pi G f^2 \hx^r_1\hx^t_2}{\sqrt h} \left({p^r}_c
   p^{rc} -\frac{1}{n-2} p^{rr} p
\right), \nonumber \\
D&=&f^2 \hx^r_1 D_c (\hx^c_2 p^{rr}) -f^2 \hx^r_1 D_c (\hx^r_2
p^{rc}) -f^2 \hx^r_1 D_c (\hx^r_2 p^{rc} ).
\eeq
Term $B$ reduces to
\beq
B
=-(16 \pi G)^{-1}f^3 \hx^r_1
 (h^{rr})^2 \pa_r \hx^t_2 \pa_r \sqrt{\si} \leq O(N).
\eeq
In order to compute the term $A$, one needs to compute the curvature
tensor.
\beq
\label{Rrr(App)}
R_{rr}
&=&-\pa_r \G^\al_{\al r} +\G^r_{rr} \G^\al_{r \al} +O(N^{-2}) =O(N^{-3}),\no\\
R_{\al \be}
&=&\pa_r \G^r_{\al \be} +\G^r_{\al \be} \G^r_{rr} +O(1) =O(N^{-1}),\no\\
R&=&h^{rr} R_{rr} +h^{\al \be} R_{\al \be} =O(N^{-1}).
\eeq
From this result, it is easy to see that
\beq
A=-\f{f^3}{16 \pi G} \hx^r_1\hx^t_2 \sqrt{\si} \left((h^{rr})^2 R_{rr}-\frac{1}{2}
h^{rr} R\right) \leq O(N).
\eeq
Term $D$ reduces to
\beq
\label{4all(App)}
D
=-f^2 \hx^r_1 D_c (\hx^c_2 p^{rr}) +2f^2 \hx^r_1 D_\al (\hx^r_2
p^{r \al} ).
\eeq

The first term of this equation becomes
\beq
-f^2 \hx^r_1 \sqrt{h}^{-1} \pa_r (\sqrt{h} \hx^r_2) p^{rr}
+f^2 \hx^r_1 \hx^r_2( \pa_r p^{rr}+2 \G^r_{rr} p^{rr} +2 \G^r_{r
  \al} p^{\al r})\leq O(N).
\eeq

On the other hand, the second term of (\ref{4all(App)}) becomes
\beq
\label{4second(App)}
&&2f^2 \hx^r_1 h^{rr} h^{\al \be} D_\al (\hx^{r}_{2} p_{r \be} )\no\\
&&=2f^2 \hx^r_1 h^{rr} h^{\al \be} [\pa_\al (\hx^{r}_{2} p_{r \be} )
+\G^r_{\alpha r} \hx^r_2 p_{r \be}-\G^a_{\al r} \hx^r_2 p_{a \be}
-\G^a_{\al \be} \hx^r_2 p_{ra} ].
\eeq
Now, from the computation of each term inside the bracket [~],
\beq
\pa_\al (\hx^r_2 p_{r \be} ) \leq O(1),~~
\G^r_{\al r} \hx^r_2 p_{r \be}
\leq  O(1), ~~
\G^a_{\al r} \hx^r_2 p_{a \be}
\leq  O(N^{-1}), ~~
\G^a_{\al \be} \hx^r_2 p_{r a}
\leq  O(N^{-1}),
\eeq
(\ref{4second(App)}) becomes
\beq
\label{B11}
2f^2 \hx^r_1 h^{rr} h^{\al \be} D_\al (\hx^{r}_{2} p_{r \be} )\leq
O(N).
\eeq
This result is contrary to the $naive$ expectation
$D_{\al}\sim\pa_{\al}$ such that this is $O(N^3)$: This implies that
$\pa_r$, which is hidden in $D_\al$ through $\G^a_{\al r}$, makes $\pa_r
O(N^3)=O(N)$.

$C$ is the most important term in (\ref{delpiall(App)}).

The first term of $C$ reduces to
\beq
\label{3first(App)}
\frac{8 \pi G \hx^r_1 \hx^t_2}{\sqrt{h}} \left[
p_{\al \be} p^{\al \be}-\frac{1}{n-2}
{p^\al}_\al p^\be_\be +O(N^{-3}) \right] \leq O(1).
\eeq

The second term reduces
\beq
-\frac{32 \pi G \hx^r_1 \hx^t_2}{\sqrt{h}}  \left[-\frac{1}{n-2}
{p^r}_r {p^\al}_\al +O(N^{-2})\right] \leq O(N).
\eeq
[Notice that the $n=2$ is meaningless in this formula, a separate
consideration is required for that case.]

Therefore, $C$ becomes
\beq
C&=&\frac{8 \pi G \hx^r_1 \hx^t_2}{\sqrt{h}} \left(
p_{\al \be} p^{\al \be}-\frac{1}{n-2}
{p^\al}_\al {p^\be}_\be \right)+ O(N) \leq O(1).
\eeq

On the other hand, since the Hamiltonian constraint ${\cal H}_t$
becomes, near the horizon,
\beq
{\cal H}_t
=\frac{16 \pi G}{\sqrt{h}} \left(
p_{\al \be} p^{\al \be}-\frac{1}{n-2}
{p^\al}_\al {p^\be}_\be \right)+O(N^{-2}) \leq O(N^{-3})
\eeq
from (\ref{Rrr(App)}) and (\ref{3first(App)}), $C$ is nothing but
\beq
C=\frac{\hx^r_1 \hx^t_2}{2 } {\cal H}_t +O(N) \leq O(1).
\eeq
In summary, one has
\beq
2 \oint _{r_+} d^{n-2} ~\hat{\xi_1}^r \delta_2 {p^r}_r
&=&2 \oint _{r_+} d^{n-2}~ \left(A+B+C+D\right) \no\\
&=&\oint _{r_+} d^{n-2}~\hx^r_1 \hx^t_2
{\cal H}_t +O(N),\\
A, B, D &\leq& O(N).
\eeq
Notice that I have $not$ used any of the constraint equations ${\cal
  H}_t\approx 0$, ${\cal H}_a \approx 0$.

\end{document}